\documentclass[10pt,english,preprint,floatfix]{revtex4}
\usepackage{amssymb}

\makeatletter

\usepackage{graphicx}

\newcommand{\mathbs}[1]
{\mbox{\boldmath $#1$}}
\newcommand{\mathcs}[1]
{\mbox{\textsc{#1}}}
\newcommand {\sups}[1]{\raisebox{1ex}{\small #1}}

\usepackage{babel}
\makeatother
\begin{document}

\title{Time--domain chirally--sensitive three--pulse coherent probes of
vibrational excitons in proteins}

\author{Darius Abramavicius and Shaul Mukamel}

\begin{abstract}
The third order optical response of bosonic excitons is calculated
using the Green's function solution of the Nonlinear Exciton
Equations (NEE) which establish a quasiparticle-scattering mechanism
for optical nonlinearities. Both time ordered and non ordered forms of the response function which represent time and frequency domain techniques, respectively, are derived. New components of the
response tensor are predicted for isotropic ensembles of periodic chiral structures to first order in the optical
wavevector. The
nonlocal nonlinear response function is calculated in momentum
space, where the finite exciton-exciton interaction length greatly reduces
the computational effort. Applications are made to coupled
anharmonic vibrations in the amide I infrared band of peptides.
Chirally--sensitive and non sensitive signals for
$\alpha$ helices and antiparallel $\beta$ sheets are compared.
\end{abstract}

\affiliation{Chemistry department, University of California, Irvine, California 92697-2025 }

\date{\today{}}

\preprint{\emph{Submitted to Chemical Physics}}

\maketitle

\section{Introduction}

Calculating the nonlinear optical response of large molecules using
conventional sum-over-states expressions is a challenging task since
the number of states accessible by multiple quantum excitations
increases rapidly with system size \cite{Mukamel_inzyss1993,mukbook}. For example, the Frenkel exciton
model for $\mathcal{M}$ coupled three level systems has
$\mathcal{M}$ one-exciton states and
$\mathcal{M}\times(\mathcal{M}+1)/2$ two-exciton states. 
Diagonalizing the two-exciton hamiltonian is the bottleneck in 
numerical simulations. The Nonlinear Exciton Equations (NEE) \cite{leegwaterMukamel92,CherniakMukamel96,KuhnCherniak96} offer
an alternative exact method for computing the exact nonlinear optical
response
of systems whose Hamiltonian conserve the number of excitons. The NEE establish an exciton
scattering mechanism and provide a practical algorithm for
computing the third order optical response, totally avoiding the
calculation of two-exciton states. The NEE were first developed
by Spano and Mukamel \cite{spanomuk89,spanomuk91a,spanomuk91b} and
applied for four-wave mixing of coupled two-level
\cite{leegwaterMukamel92,BernarzKnoester01,abramaviciusmukamel04jpcLHC2}
and three-level
\cite{KuhnCherniak96,PiryatinskiMukamelchemphys2001,scheurer01,Mukamel_inzyss1993}
molecules. The local-field-approximation was generalized by
adding two-exciton variables which properly account for double
quantum resonances. Additional dynamical variables have subsequently
been added, allowing to describe population transport via the
Redfield equation for the reduced exciton density matrix
\cite{Chernyak_physrep1995,KnoesterMukamel1991}. The NEE were
further extended to particles with arbitrary commutation relations,
and to Wannier excitons in semiconductors
\cite{CherniakMukamel96,ChernyakMukamel1996,AxtMukamel1998rev}. 
We have recently studied the frequency-domain third order
susceptibility of isotropic exciton systems and calculated the
leading terms of its tensor elements going beyond the dipole approximation
\cite{Abramavicius2005jcp-chiral}. In this paper we apply these
results to the time domain response function in a collinear field
configuration, which leads to the strongest signals.

The NEE provide a unified treatment of vibrational and
electronic excitons. The
Frenkel excitons corresponding to electronic excitations are Paulions whose non-Boson operator statistics results in
nonlinear hard-core repulsive interactions between excitons. The NEE
were originally derived using a Bosonization procedure
\cite{Dyson1956,agranovichtosich,IlinskaiaIlinski1996,spanomuk89,spanomuk91a,spanomuk91b,Chernyak_physrep1995}
 whereby the
Pauli exciton operators are replaced by Boson operators and a
repulsive potential is added to the hamiltonian. 
This results in a soft-core boson model with a finite 
anharmonicity $\Delta$; the original hard-core boson  hamiltonian is recovered by
setting $\Delta\rightarrow\infty$.

This article focuses on vibrational excitons which are
intrinsically bosons \cite{PiryatinskiMukamelchemphys2001,scheurer01,Abramavicius2005jcp-chiral}
and the finite anharmonicities serve as the source of nonlinearities.
Third order impulsive optical techniques performed with linearly
polarized light (figure \ref{enu:Scheme}) are commonly
used to probe ultrafast processes in isotropic systems. The
corresponding nonlinear response function $S_{\nu_{4}\nu_{3}\nu_{2}\nu_{1}}^{(3)}$
is a fourth rank tensor \cite{MukamelAbramavicius2004} which relates
the third order polarization, $\mathbs{P}^{(3)}(\mathbf{r},t)$, at
position $\mathbf{r}$ and time $t$ to the optical field
$\mathbs{E}(\mathbf{r}',t')$; $\nu_{i}=x,y,z$ are the polarization
components of the $i$'th laser pulse  in the lab frame. In general
the response function has 3\sups{4} tensor elements. For isotropic
systems in the dipole approximation, only those elements with
$\delta_{\nu_{4}\nu_{3}}\delta_{\nu_{2}\nu_{1}}$,
$\delta_{\nu_{4}\nu_{2}}\delta_{\nu_{3}\nu_{1}}$ and
$\delta_{\nu_{4}\nu_{1}}\delta_{\nu_{3}\nu_{2}}$ (e.g. $xxxx$, and
$zzyy$) survive rotational averaging, and three linearly-independent elements $xxyy$, $xyxy$ and $xyyx$, are necessary to
describe the response for an arbitrary pulse polarization configuration
\cite{andrewsthirumanchandran77jcpROT,Barron-CD-book-1982}. 
However, other elements with
an odd number of repeating indices (such as $xxxy$, $zzxy$), which
vanish in this approximation due to isotropic symmetry, appear
when the dipole approximation is relaxed. We have
recently shown that these elements are chirally-sensitive i e  appear
only in chiral molecules (``handed'' systems which are distinct from
their mirror images \cite{Barron-CD-book-1982,hicks-chirality2002}),
and vanish in racemates, equal mixtures of molecules with opposite
sense of chirality.

The dipole approximation implies that the optical electric field is
uniform across the molecule, thus, its phase factor
$\exp(i\mathbf{kr})$, where $\mathbf{k}$ is the wavevector,
does not affect the response and can be set to zero. However, the
wavevector does play an important role in the spectroscopy of chiral
molecules, where the variation of the phase induces new tensor
components of the response function to first order in the
wavevector, which vanish in the dipole approximation. The various tensor
elements can be probed
directly using different time-domain techniques which control
the sequences of optical interactions.

The difference in absorption of left- and
right-handed circularly polarized light
\cite{moscowitz62CD,tinoco62CD,Barron-CD-book-1982,woody-CD-book-1994} known as circular Dichroism (CD) is the simplest example of a wavevector--induced signal and is
related to the $S_{xy}^{(1)}$ elements of the linear
response tensor (when the field propagates along $z$). The technique is sensitive
to molecular structure and is extensively used for protein structure
determination both in the UV (180 - 220 nm $n-\pi*$ and $\pi-\pi*$
transitions) and the IR (1000 - 3500 cm\sups{-1} which covers most
of the amide vibrational bands)
\cite{freedmannafie94CD,cheesemafrisch96CD,stephensashvar96CD,boursopkova97CD,stephensdevlin00CD}.
Pattern-recognition and decomposition algorithms are used to
distinguish between $\alpha$-helical and $\beta$-sheet formations
using electronic
\cite{sreeramavenyaminov99ECD,sreeramavenyaminov00analbiochemECD,sreeramawoody00analbiochemECD}
and vibrational CD \cite{baumrukpancoska96VCD}.

Polypeptides often have almost translationally-invariant secondary
structures (helices, sheets, strands) which form periodic arrays of
localized vibrations. Analyzing secondary structure motifs, thus,
provides specific information which can be used
for studying globular proteins with different secondary
structures. Periodicity can be used to greatly reduce the
computational effort, as is the case for electronic excitations in
molecular crystals and semiconductor superlattices
\cite{scholes04,KovalevskijScholes2004,ButovChemla2002_semi,Chemla2001_semi_rev,AlbrechtGobel1996_semi,Ogawa_semi2004,AxtMukamel1998rev}.
Due to different translational properties of one-exciton and
two-exciton states in the Frenkel exciton model, multi-exciton
resonances cannot be generally calculated analytically even for
infinite periodic systems. The NEE only require
the one-exciton states, and yield closed expressions for infinite
periodic systems, where translational symmetry helps reduce the
problem size even further.
Relaxing the dipole approximation using the NEE, is straightforward.
The CD spectra of molecular aggregates  modelled as
 collections of coupled electric dipoles were calculated to first
order in wavevectors \cite{wagersreitermukamel}.  This model has been
applied to biological light harvesting antennae and cylindrical aggregates
\cite{somsengrondelle96CD,didragaklugkist02CD}. By extending this
procedure to the nonlinear response we obtain the complete set
of tensor elements for the response function of infinite periodic
systems.

In section \ref{sec:NonLinearAbsorption} we present the time-domain
expressions for the third order optical response. The hamiltonian
and the NEE for vibrational excitons are given in section
\ref{sec:The-NEE-for-vibrational-excitons} and the third order
response function is derived in section
\ref{sec:The-nonlinear-optical-response}. Two techniques for probing
one-exciton and two-exciton resonances are discussed in section
\ref{sec:Application-to-peptides}. The two-dimensional
infrared frequency correlation signals of two typical
structural motifs of polypeptides: the one-dimensional $\alpha$
helix and the two-dimensional antiparallel $\beta$ sheet (figure
\ref{enu:Scheme})
in the amide I region (1550 - 1750 cm\sups{-1}) are compared.The results are discussed in section
\ref{sec:Discussion}. The linear absorption is calculated in
appendix \ref{sec:LASignal}.
Rotational averages for isotropic systems are
given in appendix \ref{sec:RotationalAveraging}.
The nonlinear signals are calculated in
appendices \ref{sec:TDSignals} (four-wave mixing in various phase matching
directions) and \ref{sec:PPSignal} (pump-probe). The frequency domain signal is given in appendix \ref{sec:susceptibilities} and the exciton scattering
matrix for an infinite periodic system is presented in appendix
\ref{sec:The-exciton-infinite}.

\section{Time--domain optical response of excitons}

\label{sec:NonLinearAbsorption}

The optical response of molecules is determined by the induced
polarization, which serves as the source in the Maxwell equations
for the generated signal field. The response functions
$S_{\nu_{n+1}\nu_{n}...\nu_{1}}^{(n)}$ are system property-tensors
which allow to calculate the induced polarization for an arbitrary
incoming pulse configuration \cite{mukbook}:
\begin{eqnarray}
 &  & \mathbs{P}_{\nu_{n+1}}^{(n)}(\mathbf{x}_{n+1})=\sum_{\nu_{n},...,\nu_{1}}\int d\mathbf{x}_{n}...\int d\mathbf{x}_{1}\nonumber \\
 &  & \quad\quad S_{\nu_{n+1}\nu_{n}...\nu_{1}}^{(n)}(\mathbf{x}_{n+1},\mathbf{x}_{n}...\mathbf{x}_{1})\mathbs{E}_{\nu_{n}}(\mathbf{x}_{n})...\mathbs{E}_{\nu_{1}}(\mathbf{x}_{1}).\label{eq:NonlinearResponse}
\end{eqnarray}
Here $\mathbs{E}$ is the optical electric field vector and $\nu=x,y,z$
denote its Cartesian components in the lab frame;
$\mathbf{x}=(\mathbf{r},t)$ is the space-time vector and $\int
d\mathbf{x}\equiv\int d\mathbf{r}\int dt$, where the $\mathbf{r}$
integration runs over the molecular volume, while the $t$
integration is from $-\infty$ to $+\infty$. $\mathbf{x}$
represent the times and coordinates of the interaction with the
optical pulses. 

Using causality $S^{(n)}$ vanishes unless
$t_{1}$ \ldots{}$t_{n}$ precede $t_{n+1}$ for all $n$. 
We focus on time-domain
experiments where the interaction sequence is controlled by
short optical pulses and the
time-ordered response function is finite for
$t_{n+1}>t_{n}>...>t_{2}>t_{1}$ and vanish otherwise as shown in
figure \ref{enu:Notations}a. Alternatively we can require that the response function to be symmetric with respect to permutation of $\nu_j \mathbf{x}_j$, $j=1,2,3$. This non-time-ordered response function, which is useful for frequency domain techniques is given in appendix \ref{sec:susceptibilities}. 

The linear polarization is given by:
\begin{eqnarray}
 &  & \mathbs{P}_{\nu_{2}}^{(1)}(\mathbf{x}_{2})=\sum_{\nu_{1}}\int d\mathbf{x}_{1}S_{\nu_{2}\nu_{1}}^{(1)}(\mathbf{x}_{2},\mathbf{x}_{1})\mathbs{E}_{\nu_{1}}(\mathbf{x}_{1}),\label{eq:LinearResponse}
\end{eqnarray}
 where $S_{\nu_{2}\nu_{1}}^{(1)}(\mathbf{x}_{2},\mathbf{x}_{1})$
is the nonlocal linear response function. $S^{(1)}$ is responsible for
linear absorption (local response; see appendix \ref{sec:LASignal}) and CD (non local response).
Four wave mixing (4WM) signals are described by the third order polarization
\begin{eqnarray}
 &  & \mathbs{P}_{\nu_{4}}^{(3)}(\mathbf{x}_{4})=\sum_{\nu_{3}\nu_{2}\nu_{1}}\int d\mathbf{x}_{3}\int d\mathbf{x}_{2}\int d\mathbf{x}_{1}\nonumber \\
 &  & \quad\quad S_{\nu_{4}\nu_{3}\nu_{2}\nu_{1}}^{(3)}(\mathbf{x}_{4},\mathbf{x}_{3},\mathbf{x}_{2},\mathbf{x}_{1})\mathbs{E}_{\nu_{3}}(\mathbf{x}_{3})\mathbs{E}_{\nu_{3}}(\mathbf{x}_{2})\mathbs{E}_{\nu_{1}}(\mathbf{x}_{1}).\label{eq:ThirdResponse}
\end{eqnarray}
We assume three short well-separated incoming pulses:
\begin{equation}
\mathbs{E}_{\nu}(\mathbf{x})=\frac{1}{2}\sum_{s=1}^{3}\tilde{E}_{\nu}^{(s)}(t-t_{s})\exp(i\mathbf{k}_{s}\mathbf{r}-i\omega_{s}(t-t_{s})+i\phi_{s})+c.c.,\label{eq:ActingOptField}
\end{equation}
where $\tilde{E}_{\nu}^{(s)}(t-t_{s})$ is the (real) envelope of pulse
$s$ centered at $t_{s}$, with wavevector $\mathbf{k}_{s}$, carrier
frequency $\omega_{s}$ and phase $\phi_{s}$. When the pulses
 are much shorter than the relevant molecular
timescale, 
 $\tilde{E}_{\nu}(t-t_{s})$ in eq.
(2) can be approximated as $E_{\nu}^{(s)}\delta(t-t_{s})$.
Since the pulses are longer than the optical periods we must then invoke the
rotating wave approximation - RWA and only
retain $S^{(3)}$ terms resonant with optical fields which fall within the
pulse bandwidth $\bar{\omega}_{s}$ \cite{DreyerMukamel2003}. In this approximation the field amplitude in the
frequency domain is taken to have a rectangular shape centered at
$\omega_{s}$ with width $\bar{\omega}_{s}$. We next assume that the
pulses are temporally well separated and ordered, i. e. the first pulse
$\tilde{E}^{(1)}$ interacts at $t_{1}$, followed by $\tilde{E}^{(2)}$
at $t_{2}$ and the last pulse is $\tilde{E}^{(3)}$ at
$t_{3}$. The third order polarization is then given by
\begin{eqnarray}
 &  & \mathbs{P}_{\nu_{4}}^{(3)}(\bar{\mathbf{x}}_{4})=\exp(\pm i\phi_{3}\pm i\phi_{2}\pm i\phi_{1})\nonumber \\
 &  & \quad\quad\times\frac{1}{2^{3}}\sum_{\nu_{3}\nu_{2}\nu_{1}}S_{\nu_{4}\nu_{3}\nu_{2}\nu_{1}}^{(3)}(\mathbf{k}_{4}t_{4},\pm\mathbf{k}_{3}t_{3},\pm\mathbf{k}_{2}t_{2},\pm\mathbf{k}_{1}t_{1})E_{\nu_{3}}^{(3)}E_{\nu_{2}}^{(2)}E_{\nu_{1}}^{(1)},\label{eq:ThirdResponseRWA}
 \end{eqnarray}
where $\bar{\mathbf{x}}\equiv(\mathbf{k},t)$; $t_{j}$ and
$\mathbf{k}_{j}$ now coincide with the central pulse times and
wavevectors and we have applied the space-time Fourier transform
   $
F(\mathbf{k},\omega)=\int dt\int d\mathbf{r}\exp(i\omega t+i\mathbf{kr})F(\mathbf{r},t)
   $.

When the system is initially at equilibrium, the
response function is time translationally invariant
and only depends on the positive time intervals $T_{s} \equiv
t_{s+1}-t_{s}$ (see figure \ref{enu:Notations}). Space translational invariance for
isotropic systems implies $\mathbf{k}_{4}=\mp\mathbf{k}_{3}\mp\mathbf{k}_{2}\mp\mathbf{k}_{1}$.
We thus denote $\mathbs{P}_{\nu_{4}}^{(3)}(-\mathbf{k}_{4},t_{4})\equiv\mathbs{P}_{\nu_{4}}^{\mathbf{k}_{S}}(T_3,T_2,T_1)$ with   $\mathbf{k}_{S}\equiv-\mathbf{k}_{4}$. There
are four independent signal wavevectors $\mathbf{k}_{S}$:
$\mathbf{k}_{I}=-\mathbf{k}_{1}+\mathbf{k}_{2}+\mathbf{k}_{3}$,
$\mathbf{k}_{II}=+\mathbf{k}_{1}-\mathbf{k}_{2}+\mathbf{k}_{3}$,
$\mathbf{k}_{III}=+\mathbf{k}_{1}+\mathbf{k}_{2}-\mathbf{k}_{3}$ and
$\mathbf{k}_{IV}=+\mathbf{k}_{1}+\mathbf{k}_{2}+\mathbf{k}_{3}$.

\section{The NEE for vibrational excitons}

\label{sec:The-NEE-for-vibrational-excitons}

The amide vibrations of polypeptides can be modeled as $\mathcal{N}$
coupled anharmonic vibrational modes localized at the peptide bonds
and described by the exciton hamiltonian:
\begin{eqnarray}
 &  & \hat{H}=\sum_{m}\varepsilon_{m}\hat{B}_{m}^{\dagger}\hat{B}_{m}+\sum_{mn}^{m\neq n}J_{m,n}\hat{B}_{m}^{\dagger}\hat{B}_{n}\nonumber \\
 &  & \quad\quad+\sum_{mn,m'n'}U_{mn,m'n'}\hat{B}_{m}^{\dagger}\hat{B}_{n}^{\dagger}\hat{B}_{m'}\hat{B}_{n'}-\int d\mathbf{r}\hat{\mathbf{P}}(\mathbf{r})\cdot\mathbs{E}(\mathbf{r},\tau).\label{eq:HamMol}
\end{eqnarray}
 The creation, $\hat{B}_{m}^{\dagger}$, and annihilation, $\hat{B}_{m}$,
operators for mode $m$ satisfy the boson
$[\hat{B}_{m},\hat{B}_{n}^{\dagger}]=\delta_{mn}$ commutation
relations. The first two terms represent the free-boson harmonic
hamiltonian where $\varepsilon_{m}$ is the frequency of mode $m$,
and the quadratic intermode coupling, $J_{m,n}$, is calculated in
the Heitler-London approximation where off resonant
$\hat{B}_{m}^{\dagger}\hat{B}_{n}^{\dagger}$ and
$\hat{B}_{m}\hat{B}_{n}$ terms are neglected. The third term represents a quartic
anharmonicity. We assume a pairwise anharmonic interaction,
$U_{mn,m'n'}=\frac{\Delta_{m,n}}{4}(\delta_{mm'}\delta_{nn'}+\delta_{mn'}\delta_{nm'})$,
where $\Delta_{mm}$ is the on-site anharmonicity of the overtone
band and $\Delta_{nm}$ is the intermode anharmonicity of the
combination band. These anharmonicities constitute the
exciton-exciton scattering potential. For $\Delta=0$ the hamiltonian
describes free bosons, which is a linear system whose nonlinear
response vanishes identically \cite{Mukamel_inzyss1993}. The fourth term in the hamiltonian
represents the interaction with the optical field
$\mathbs{E}(\mathbf{r},t)$, where
\begin{equation}
\hat{\mathbf{P}}(\mathbf{r})=\sum_{m}\delta(\mathbf{r}-\mathbf{r}_{m})\mathbs{\mu}_{m}(\hat{B}_{m}^{\dagger}+\hat{B}_{m})\label{eq:OperatorPolarization}
\end{equation}
 is the polarization operator and $\mathbs{\mu}_{m}$ is the transition
dipole moment of mode $m$ located at $\mathbf{r}_{m}$; a vector
with components ($\mathbs{\mu}_{m}^{x}$,$\mathbs{\mu}_{m}^{y}$,
$\mathbs{\mu}_{m}^{z}$).

The expectation value of the polarization operator which describes
the vibrational response to the optical field will be calculated by
solving the NEE
\cite{spanomuk91a,leegwaterMukamel92,CherniakMukamel96}. This
hierarchy of equations of motion for exciton variables may be exactly
truncated order by order in the field since the molecular
hamiltonian conserves the number of excitons, and the optical field
creates or annihilates one exciton at a time. When pure-dephasing is
neglected, the only required variables for the third order response
are $B_{m}=\langle\hat{B}_{m}\rangle$ (one-exciton) and
$Y_{mn}=\langle\hat{B}_{m}\hat{B}_{n}\rangle$ (two-exciton) and the
NEE read \cite{spanomuk91a,spanomuk91b}:
\begin{equation}
-i\frac{\partial B_{m}}{\partial\tau}+\sum_{n}h_{m,n}B_{n}=\tilde{\mu}_{m}(\tau)-\sum_{l'm'n'}V_{ml'm'n'}B_{l'}^{\ast}Y_{m'n'},\label{eq:NEEBRed}
\end{equation}
\begin{equation}
-i\frac{\partial Y_{mn}}{\partial\tau}+\sum_{m'n'}(h_{mn,m'n'}^{(Y)}+V_{mn,m'n'})Y_{m'n'}=\tilde{\mu}_{m}(\tau)B_{n}+\tilde{\mu}_{n}(\tau)B_{m}.\label{eq:NEEYRed}
\end{equation}
 here $h_{m,n}=\delta_{m,n}\varepsilon_{m}+J_{m,n}(1-\delta_{m,n})$
is an effective one-exciton hamiltonian,
$h_{mn,m'n'}^{(Y)}=\delta_{m',m}h_{n,n'}+\delta_{n,n'}h_{m,m'}$ is a
free-two-exciton hamiltonian, $V_{mn,m'n'}\equiv
U_{mn,m'n'}+U_{nm,m'n'}=\frac{\Delta_{m,n}}{2}(\delta_{mm'}\delta_{nn'}+\delta_{mn'}\delta_{nm'})$
is the anharmonicity matrix and
$\tilde{\mu}_{m}(\tau)=\mathbs{\mu}_{m}\cdot\mathbs{E}(\mathbf{r}_{m},\tau)$ (for the actual times we use $\tau$ instead of $t$ which were defined as time-ordered).
The polarization is given by the expectation value of eq. (\ref{eq:OperatorPolarization}):
\begin{equation}
\mathbs{P}(\mathbf{r},\tau)=\sum_{m}\delta(\mathbf{r}-\mathbf{r}_{m})\mathbs{\mu}_{m}B_{m}(\tau)+c.c.\label{eq:PolarizationViaB}
\end{equation}
The nonlinearities in these equations originate from the
anharmonicity: as indicated earlier, for $V=0$ eq. (\ref{eq:NEEBRed}) is linear, and the
nonlinear response vanishes (the two-exciton variable can be exactly
factorized as $Y_{mn}=B_{m}B_{n}$ in eq. (\ref{eq:NEEYRed}) ).

The evolution of a single-exciton created by an impulsive
excitation, $B_{m}^{(1)}(\tau)$, is described by the one-exciton
Green's function $G(\tau)$:
\begin{equation}
B_{m}^{(1)}(\tau)=\sum_{m'}G_{m,m'}(\tau)B_{m'}^{(1)}(0),\label{eq:GBDefinition}
\end{equation}
 which satisfies the equation:
 \begin{equation}
\frac{dG_{m,n}(\tau)}{d\tau}+i\sum_{n'}h_{m,n'}G_{n',n}(\tau)=\delta(\tau).\label{eq:GBEquation}
\end{equation}
 This equation can be solved using the one-exciton eigenenergies,
$\mathcs{e}_{\xi}$, and eigenvectors, $\psi_{\xi m}$:
\begin{equation}
\sum_{n}h_{m,n}\psi_{\xi n}=\mathcs{e}_{\xi}\psi_{\xi m}.\label{eq:1ExEigenEquation}
\end{equation}
 We then get
 \begin{equation}
G_{m,n}(\tau)=\sum_{\xi}\psi_{\xi m}I_{\xi}(\tau)\psi_{\xi n}^{\ast},\label{eq:1ExGExpression}
\end{equation}
 where
 \begin{equation}
I_{\xi}(\tau)=\theta(\tau)\exp(-i\mathcs{e}_{\xi}\tau-\gamma_{\xi}\tau),\label{eq:GreensEigen}
\end{equation}
 $\gamma_{\xi}$ is a dephasing rate of the $\xi$ exciton state, and
the Heavyside step function ($\theta(\tau)=0$ for $\tau<0$ and
$\theta(\tau)=1$ for $\tau\geq0$) guarantees causality.

Similarly we define the two-exciton
Green's function $\mathcal{G}^{Y}$:
\begin{equation}
Y_{mn}(\tau)=\sum_{m'n'}\mathcal{G}_{mn,m'n'}^{Y}(\tau)Y_{m'n'}(0),\label{eq:GYDefinition}
\end{equation}
whose equation of motion:
 \begin{equation}
\frac{d\mathcal{G}_{mn,m'n'}^{Y}}{dt}+i\sum_{m''n''}(h_{mn,m''n''}^{(Y)}+V_{mn,m''n''})\mathcal{G}_{m''n'',m'n'}^{Y}=\delta(\tau).\label{eq:GYEquation}
\end{equation}
 $\mathcal{G}^{Y}$ is connected to the zero-order noninteracting
($V=0$) two-exciton Green's function $\mathcal{G}$ by the
Bethe-Salpeter equation
\cite{leegwaterMukamel92,CherniakMukamel96}:
\begin{equation}
\mathcal{G}^{Y}(\tau)=\mathcal{G}(\tau)+\int_{0}^{\tau}d\tau'\int_{0}^{\tau'}d\tau_{1}\mathcal{G}(\tau-\tau')\Gamma(\tau'-\tau_{1})\mathcal{G}(\tau_{1}),\label{eq:BetheSalpeterTime}
\end{equation}
 which defines the \emph{two exciton scattering matrix} $\Gamma$.
Both the two-exciton Green's function and the scattering matrix are tetradic
matrices; like $G(\tau)$, the scattering matrix is causal as well.
$\mathcal{G}$ can be factorized into a product of
one-exciton Green's functions,
$\mathcal{G}_{mn,m'n'}(\tau)=G_{m,m'}(\tau)G_{n,n'}(\tau)$.

The frequency domain scattering matrix $\Gamma(\omega)$ obtained by
the Fourier transform,    $
\Gamma(\omega)=\int dt\exp(i\omega t) \Gamma(t)
   $, is calculated in appendix \ref{sec:The-exciton-infinite}
 \cite{leegwaterMukamel92,Abramavicius2005jcp-chiral}:
\begin{equation}
\Gamma(\omega)=-iV(1+i\mathcal{G}(\omega)V)^{-1}.\label{eq:ScattMatrixW}
\end{equation}
 where
 \begin{equation}
\mathcal{G}_{mn,m'n'}(\omega)=\sum_{\xi\xi'}\psi_{\xi m}\psi_{\xi'n}\mathcal{I}_{\xi\xi'}(\omega)\psi_{\xi m'}^{\ast}\psi_{\xi'n'}^{\ast},\label{eq:GY0ExpressionTime}
\end{equation}
 with\begin{equation}
\mathcal{I}_{\xi\xi'}(\omega)=\frac{i}{\omega-\mathcs{e}_{\xi}-\mathcs{e}_{\xi'}+i(\gamma_{\xi}+\gamma_{\xi'})}.\label{eq:2ExcGreenFree}
\end{equation}
 Calculating the scattering matrix requires the inversion of
the matrix $D=1+i\mathcal{G}(\omega)V$ whose matrix elements:
\begin{equation}
D_{mn,ij}(\omega)=\delta_{mi}\delta_{nj}+i\sum_{m'n'}\mathcal{G}_{mn,m'n'}(\omega)V_{m'n',ij}.\label{eqApp:DeltaExpanded}
\end{equation}
The required numerical effort can be reduced considerably for periodic systems and when the short
range nature of anharmonicities is taken into account \cite{leegwaterMukamel92}.

\section{The nonlinear optical response: Green's function solution of the
NEE}

\label{sec:The-nonlinear-optical-response}

The nonlinear response is calculated by an order-by-order expansion
of the NEE variables in the field using the exciton Green's functions,
where the optical field and the lower order variables serve as the
sources. Setting $B_{m}=B_{m}^{(1)}+B_{m}^{(2)}+B_{m}^{(3)}+...$
and $Y_{mn}=Y_{mn}^{(1)}+Y_{mn}^{(2)}+Y_{mn}^{(3)}+...$ we get to
third order \cite{CherniakMukamel96,Abramavicius2005jcp-chiral}:
\begin{eqnarray}
 &  & B_{n_{4}}^{(3)}(\tau_{4})=\nonumber \\
 &  & \quad\quad2i\int_{-\infty}^{\infty}d\tau''\int_{-\infty}^{\infty}d\tau'\int_{-\infty}^{\infty}d\tau_{3}\int_{-\infty}^{\infty}d\tau_{2}\int_{-\infty}^{\infty}d\tau_{1}\nonumber \\
 &  & \quad\quad\sum_{n_{1}n_{2}n_{3}}\sum_{n_{1}'n_{2}'n_{3}'n_{4}'}\theta(\tau_{2}-\tau_{1})\Gamma_{n_{4}'n_{3}',n_{2}'n_{1}'}(\tau''-\tau')\nonumber \\
 &  & \quad\quad\times G_{n_{4},n_{4}'}(\tau_{4}-\tau'')G_{n_{3}',n_{3}}^{\dagger}(\tau''-\tau_{3})G_{n_{2}',n_{2}}(\tau'-\tau_{2})G_{n_{1}',n_{1}}(\tau'-\tau_{1})\nonumber \\
 &  & \quad\quad\times\tilde{\mu}_{n_{3}}(\tau_{3})\tilde{\mu}_{n_{2}}(\tau_{2})\tilde{\mu}_{n_{1}}(\tau_{1}),\label{eq:B3Complete}
 \end{eqnarray}
 The $\tau'$ and $\tau''$ variables denote the times of the first
and the last exciton-exciton interaction respectively, as shown in
figure \ref{enu:Notations}a. The third order polarization is finally
obtained from eqs. (\ref{eq:PolarizationViaB}) and (\ref{eq:B3Complete}):
\begin{eqnarray}
 &  & \mathbs{P}_{\nu_{4}}^{(3)}(\mathbf{r}_{4}\tau_4)=2i\int d\mathbf{r}_{3}\int d\mathbf{r}_{2}\int d\mathbf{r}_{1}\int_{-\infty}^{\infty}d\tau_{3}\int_{-\infty}^{\infty}d\tau_{2}\int_{-\infty}^{\infty}d\tau_{1}\nonumber \\
 &  & \quad\quad\sum_{n_{4}n_{3}n_{2}n_{1}}\langle\mathbf{M}_{n_{4}n_{3}n_{2}n_{1}}^{\nu_{4}\nu_{3}\nu_{2}\nu_{1}}(\mathbf{r}_{4},\mathbf{r}_{3},\mathbf{r}_{2},\mathbf{r}_{1})\rangle\nonumber \\
 &  & \quad\quad\times\int_{-\infty}^{\infty}d\tau''\int_{-\infty}^{\infty}d\tau'\sum_{n_{1}'n_{2}'n_{3}'n_{4}'}\theta(\tau_{2}-\tau_{1})\Gamma_{n_{4}'n_{3}',n_{2}'n_{1}'}(\tau''-\tau')\nonumber \\
 &  & \quad\quad\times G_{n_{4},n_{4}'}(\tau_{4}-\tau'')G_{n_{3}',n_{3}}^{\dagger}(\tau''-\tau_{3})G_{n_{2}',n_{2}}(\tau'-\tau_{2})G_{n_{1}',n_{1}}(\tau'-\tau_{1})\nonumber \\
 &  & \quad\quad\times 
 \mathbs{E}_{\nu_{3}}(\mathbf{r}_{3}\tau_{3})
 \mathbs{E}_{\nu_{2}}(\mathbf{r}_{2}\tau_{2})
 \mathbs{E}_{\nu_{1}}(\mathbf{r}_{1}\tau_{1})+c.c.,\label{eq:P3Complete}
 \end{eqnarray}
 where
 \begin{eqnarray}
 &  & \mathbf{M}_{n_{4}n_{3}n_{2}n_{1}}^{\nu_{4}\nu_{3}\nu_{2}\nu_{1}}(\mathbf{r}_{4},\mathbf{r}_{3},\mathbf{r}_{2},\mathbf{r}_{1})=\delta(\mathbf{r}_{4}-\mathbf{r}_{n_{4}})\delta(\mathbf{r}_{3}-\mathbf{r}_{n_{3}})\delta(\mathbf{r}_{2}-\mathbf{r}_{n_{2}})\delta(\mathbf{r}_{1}-\mathbf{r}_{n_{1}})\nonumber \\
 &  & \qquad\times\mathbs{\mu}_{n_{4}}^{\nu_{4}}\mathbs{\mu}_{n_{3}}^{\nu_{3}}\mathbs{\mu}_{n_{2}}^{\nu_{2}}\mathbs{\mu}_{n_{1}}^{\nu_{1}},\label{eq:Mdefinition}
 \end{eqnarray}
 ``c.c'' is the complex conjugate and $\langle\ldots\rangle$
denotes rotational averaging (appendix
\ref{sec:RotationalAveraging}). It is important to note that unlike
$t_{1}$, $t_{2}$ and $t_{3}$, the integration variables $\tau_{1}$,
$\tau_{2}$ and $\tau_{3}$ do not have any particular time ordering.

Since the Green's functions are retarded
(i. e. they vanish for negative time arguments) only three sequences of interaction times contribute to eq.
(\ref{eq:P3Complete}): i) $\tau_{2}>\tau_{1}>\tau_{3}$ , ii)
$\tau_{2}>\tau_{3}>\tau_{1}$, iii) $\tau_{3}>\tau_{2}>\tau_{1}$. In
each case we switch to a different set of \emph{time-ordered} variables: i)
$t_{4}=\tau_{4}$, $t_{3}=\tau_{2}$, $t_{2}=\tau_{1}$,
$t_{1}=\tau_{3}$, ii) $t_{4}=\tau_{4}$, $t_{3}=\tau_{2}$,
$t_{2}=\tau_{3}$, $t_{1}=\tau_{1}$, iii) $t_{4}=\tau_{4}$,
$t_{3}=\tau_{3}$, $t_{2}=\tau_{2}$, $t_{1}=\tau_{1}$ (figure
\ref{enu:Notations}b). Eq. (\ref{eq:ThirdResponse}) together with eq. (\ref{eq:P3Complete}) then result in the following three contributions to the
response function:
\begin{eqnarray}
 &  & S_{\nu_{4}...\nu_{1}}^{(3)}(\mathbf{x}_{4},...,\mathbf{x}_{1})=S_{\nu_{4}...\nu_{1}}^{(I)}(\mathbf{x}_{4},...,\mathbf{x}_{1})\nonumber \\
 &  & \quad\quad+S_{\nu_{4}...\nu_{1}}^{(II)}(\mathbf{x}_{4},...,\mathbf{x}_{1})\nonumber \\
 &  & \quad\quad+S_{\nu_{4}...\nu_{1}}^{(III)}(\mathbf{x}_{4},...,\mathbf{x}_{1})+c.c.\label{eq:TimeOrderedResponse}
 \end{eqnarray}
Each term now corresponds to one particular
interaction sequence ( $S^{(I)}$ is obtained from (i),
  $S^{(II)}$ -- from (ii) and  $S^{(III)}$ -- from (iii) ).
These are given by:
 \begin{eqnarray}
 &  & S_{\nu_{4}...\nu_{1}}^{(I)}(\mathbf{x}_{4},...,\mathbf{x}_{1})=2i\sum_{n_{4}...n_{1}}\langle\mathbf{M}_{n_{4}n_{3}n_{2}n_{1}}^{\nu_{4}\nu_{3}\nu_{2}\nu_{1}}(\mathbf{r}_{4},\mathbf{r}_{3},\mathbf{r}_{2},\mathbf{r}_{1})\rangle\nonumber \\
 &  & \quad\quad\times\int_{0}^{t_{43}}d\tau_{s}''\int_{0}^{\tau_{s}''}d\tau_{s}'\sum_{n_{4}'n_{3}'n_{2}'n_{1}'}\Gamma_{n_{4}'n_{1}'n_{3}'n_{2}'}(\tau_{s}''-\tau_{s}')\nonumber \\
 &  & \quad\quad\times G_{n_{4}n_{4}'}(\tau_{s}')G_{n_{3}'n_{3}}(t_{43}-\tau_{s}'')G_{n_{2}'n_{2}}(t_{42}-\tau_{s}'')G_{n_{1}'n_{1}}^{\dagger}(t_{41}-\tau_{s}'),\label{eq:TOrd1}
 \end{eqnarray}
\begin{eqnarray}
 &  & S_{\nu_{4}...\nu_{1}}^{(II)}(\mathbf{x}_{4},...,\mathbf{x}_{1})=2i\sum_{n_{4}...n_{1}}\langle\mathbf{M}_{n_{4}n_{3}n_{2}n_{1}}^{\nu_{4}\nu_{3}\nu_{2}\nu_{1}}(\mathbf{r}_{4},\mathbf{r}_{3},\mathbf{r}_{2},\mathbf{r}_{1})\rangle\nonumber \\
 &  & \quad\quad\times\int_{0}^{t_{43}}d\tau_{s}''\int_{0}^{\tau_{s}''}d\tau_{s}'\sum_{n_{4}'n_{3}'n_{2}'n_{1}'}\Gamma_{n_{4}'n_{2}'n_{3}'n_{1}'}(\tau_{s}''-\tau_{s}')\nonumber \\
 &  & \quad\quad\times G_{n_{4}n_{4}'}(\tau_{s}')G_{n_{3}'n_{3}}(t_{43}-\tau_{s}'')G_{n_{2}'n_{2}}^{\dagger}(t_{42}-\tau_{s}')G_{n_{1}'n_{1}}(t_{41}-\tau_{s}''),\label{eq:TOrd2}
 \end{eqnarray}
\begin{eqnarray}
 &  & S_{\nu_{4}...\nu_{1}}^{(III)}(\mathbf{x}_{4},...,\mathbf{x}_{1})=2i\sum_{n_{4}...n_{1}}\langle\mathbf{M}_{n_{4}n_{3}n_{2}n_{1}}^{\nu_{4}\nu_{3}\nu_{2}\nu_{1}}(\mathbf{r}_{4},\mathbf{r}_{3},\mathbf{r}_{2},\mathbf{r}_{1})\rangle\nonumber \\
 &  & \quad\quad\times\int_{0}^{t_{42}}d\tau_{s}''\int_{0}^{\tau_{s}''}d\tau_{s}'\sum_{n_{4}'n_{3}'n_{2}'n_{1}'}\Gamma_{n_{4}'n_{3}'n_{2}'n_{1}'}(\tau_{s}''-\tau_{s}').\nonumber \\
 &  & \quad\quad\times G_{n_{4}n_{4}'}(\tau_{s}')G_{n_{3}'n_{3}}^{\dagger}(t_{43}-\tau_{s}')G_{n_{2}'n_{2}}(t_{42}-\tau_{s}'')G_{n_{1}'n_{1}}(t_{41}-\tau_{s}'')\label{eq:TOrd3}
 \end{eqnarray}
Here $\tau_{s}'$  denotes the time interval between the polarization
detection and first exciton scattering event, while $\tau_{s}''$
denotes the interval between the detection and the last exciton
scattering event, as shown in figure \ref{enu:Notations}a. Using
these time-ordered expressions we can switch to new variables
representing the time intervals between interactions $t_{ij}=t_i-t_j$ with $i>j$:
$t_{43} = T_3$, $t_{42}=T_{3}+T_{2}$, $t_{41}=T_{3}+T_{2}+T_{1}$.
Thus, all terms in the response function only depend on these three
positive time intervals. So far, the
three terms in the response function merely represent different time orderings in the integrations, however, we will
shortly see that they represent distinct optical signals.

The time intervals in eqs. (\ref{eq:TOrd1})--(\ref{eq:TOrd3}) and
their relations to the actual interaction times are depicted in
figure \ref{enu:Notations}. The response can be interpreted using
figure \ref{enu:Notations}c: Let us consider $S^{(III)}$: three
interactions with the optical fields at times $t_{1}<t_{2}<t_{3}$
generate three quasi-particles (two with positive oscillation
frequency and one -- with negative) whose evolution is described by
the one-exciton Green's functions $G_{n_{1}',n_{1}}$,
$G_{n_{2}',n_{2}}$ and $G_{n_{3}',n_{3}}^{\dagger}$. The two
positive frequency quasiparticles at $n_{1}'$ and $n_{2}'$ are
scattered by $\Gamma_{n_{4}'n_{3}'n_{2}'n_{1}'}$ to $n_{3}'$ and
$n_{4}'$. One of the two excitons, $n_{3}'$, corresponds to the
exciton generated by the third field, $n_{3}'\leftarrow n_{3}$. The
other exciton, $n_{4}'\to n_{4}$, generates the optical response.
$S^{(I)}$ and $S^{(II)}$, differ by the scattering sequence of events as shown
in figure \ref{enu:Notations}c and may be interpreted similarly.

Using eq. (\ref{eq:ThirdResponse}) and invoking the same
approximations used in section \ref{sec:NonLinearAbsorption} we can
relate the wavevectors and times in the response functions (eqs.
(\ref{eq:TOrd1})-(\ref{eq:TOrd3})
) to the wavevectors and times of the optical pulses. Assuming that
the carrier frequencies of all pulses are resonant with the
one-exciton manifold, we select the resonances in the response
functions using the RWA and express the nonlinear polarization for
the signal wavevectors $\mathbf{k}_{I}$, $\mathbf{k}_{II}$ and
$\mathbf{k}_{III}$. To find out the wavevector dependence we transform
eqs. (\ref{eq:TOrd1})-(\ref{eq:TOrd3}) to momentum space using $F(\mathbf{k}) = \int d\mathbf{r} F(\mathbf{r}) \exp(i\mathbf{kr})$. Then:
\begin{eqnarray}
 &  & \mathbs{P}_{\nu_{4}}^{\mathbf{k}_{I}}(T_{3},T_{2},T_{1})=\frac{1}{2^{3}}\exp(+i\phi_{3}+i\phi_{2}-i\phi_{1})\nonumber \\
 &  & \quad\quad\times\sum_{\nu_{3}\nu_{2}\nu_{1}}\mathbb{S}_{\nu_{4}\nu_{3}\nu_{2}\nu_{1}}^{\mathbf{k}_{I}}(T_3,T_2,T_1)E_{\nu_{3}}^{(3)}E_{\nu_{2}}^{(2)}E_{\nu_{1}}^{(1)},\label{eq:ThirdResponseRWA1}
\end{eqnarray}
\begin{eqnarray}
 &  & \mathbs{P}_{\nu_{4}}^{\mathbf{k}_{II}}(T_{3},T_{2},T_{1})=\frac{1}{2^{3}}\exp(+i\phi_{3}-i\phi_{2}+i\phi_{1})\nonumber \\
 &  & \quad\quad\times\sum_{\nu_{3}\nu_{2}\nu_{1}}\mathbb{S}_{\nu_{4}\nu_{3}\nu_{2}\nu_{1}}^{\mathbf{k}_{II}}(T_3,T_2,T_1)E_{\nu_{3}}^{(3)}E_{\nu_{2}}^{(2)}E_{\nu_{1}}^{(1)},\label{eq:ThirdResponseRWA2}
\end{eqnarray}
\begin{eqnarray}
 &  & \mathbs{P}_{\nu_{4}}^{\mathbf{k}_{III}}(T_{3},T_{2},T_{1})=\frac{1}{2^{3}}\exp(-i\phi_{3}+i\phi_{2}+i\phi_{1})\nonumber \\
 &  & \quad\quad\times\sum_{\nu_{3}\nu_{2}\nu_{1}}\mathbb{S}_{\nu_{4}\nu_{3}\nu_{2}\nu_{1}}^{\mathbf{k}_{III}}(T_3,T_2,T_1)E_{\nu_{3}}^{(3)}E_{\nu_{2}}^{(2)}E_{\nu_{1}}^{(1)},\label{eq:ThirdResponseRWA3}
\end{eqnarray}
where we have denoted
\begin{eqnarray}
\mathbb{S}_{\nu_{4}\nu_{3}\nu_{2}\nu_{1}}^{\mathbf{k}_{I}}(T_3,T_2,T_1) \equiv S_{\nu_{4}\nu_{3}\nu_{2}\nu_{1}}^{(I)}((\mathbf{k}_{1}-\mathbf{k}_{2}-\mathbf{k}_{3})t_{4},\mathbf{k}_{3}t_{3},\mathbf{k}_{2}t_{2},-\mathbf{k}_{1}t_{1}),\label{eq:ThirdResponseRWA1_def}
\end{eqnarray}
\begin{eqnarray}
\mathbb{S}_{\nu_{4}\nu_{3}\nu_{2}\nu_{1}}^{\mathbf{k}_{II}}(T_3,T_2,T_1) \equiv S_{\nu_{4}\nu_{3}\nu_{2}\nu_{1}}^{(II)}((\mathbf{k}_{2}-\mathbf{k}_{1}-\mathbf{k}_{3})t_{4},\mathbf{k}_{3}t_{3},-\mathbf{k}_{2}t_{2},\mathbf{k}_{1}t_{1}),\label{eq:ThirdResponseRWA2_def}
\end{eqnarray}
\begin{eqnarray}
\mathbb{S}_{\nu_{4}\nu_{3}\nu_{2}\nu_{1}}^{\mathbf{k}_{III}}(T_3,T_2,T_1) \equiv S_{\nu_{4}\nu_{3}\nu_{2}\nu_{1}}^{(III)}((\mathbf{k}_{3}-\mathbf{k}_{2}-\mathbf{k}_{1})t_{4},-\mathbf{k}_{3}t_{3},\mathbf{k}_{2}t_{2},\mathbf{k}_{1}t_{1}).\label{eq:ThirdResponseRWA3_def}
\end{eqnarray}
\emph{We, thus, find that the
response function $S^{(I)}$ generates the $\mathbf{k}_{I}$ signal,
$S^{(II)}$ generates $\mathbf{k}_{II}$ and $S^{(III)}$ generates
$\mathbf{k}_{III}$. } The fourth possible signal
$\mathbs{P}^{\mathbf{k}_{IV}}$ vanishes for the present model since it 
has no transition dipole connecting the three-exciton states with
the ground state. The three terms in eqs.
(\ref{eq:TOrd1})-(\ref{eq:TOrd3})
were obtained by a simple bookkeeping of time variables. The RWA has
connected these terms with the impulsive signals in eqs.
(\ref{eq:ThirdResponseRWA1})-(\ref{eq:ThirdResponseRWA3_def}). Each of
the three signals is thus given by a single term.

Equations (\ref{eq:TOrd1})-(\ref{eq:TOrd3}) are used in appendix \ref{sec:TDSignals}
to calculate the signals in the eigenstate basis. The frequency domain scattering matrix
(see appendix \ref{sec:The-exciton-infinite}) allows to simplify response functions in eqs. (\ref{eqApp:TimeOrdRespFin10})-(\ref{eqApp:TimeOrdRespFin30}).
The three signals are given by eqs.  (\ref{eq:SignalIntK1}), (\ref{eq:SignalIntK2}) and (\ref{eq:SignalIntK3}). The sequential pump-probe spectrum is calculated in appendix \ref{sec:PPSignal}.

By transforming all time variables to the frequency domain,  
$
\mathbb{S}(\Omega_3,\Omega_2,\Omega_1) = 
\int_0^{\infty}dT_3 
\int_0^{\infty}dT_2
\int_0^{\infty}dT_1
\mathbb{S}(T_3,T_2,T_1) \exp(i\Omega_3T_3+i\Omega_2T_2+i\Omega_1T_1)
$,
(see appendix \ref{sec:TDSignals}) equations (\ref{eq:ThirdResponseRWA1_def})-(\ref{eq:ThirdResponseRWA3_def})  give:
\begin{eqnarray}
 &  & \mathbb{S}_{\nu_{4}\nu_{3}\nu_{2}\nu_{1}}^{\mathbf{k}_{I}}
 (\Omega_3,\Omega_2,\Omega_1)
 =2i\sum_{\xi_{4}...\xi_{1}}
\langle\mathbs{d}_{\xi_{4}}^{\nu_{4}}(\mathbf{k}_{1}-\mathbf{k}_{2}-\mathbf{k}_{3})
        \mathbs{d}_{\xi_{3}}^{\nu_{3}\ast}(-\mathbf{k}_{3})
		\mathbs{d}_{\xi_{2}}^{\nu_{2}\ast}(-\mathbf{k}_{2})
		\mathbs{d}_{\xi_{1}}^{\nu_{1}}(-\mathbf{k}_{1})
\rangle\nonumber \\
 &  & \quad\quad\times
 I_{\xi_{1}\xi_{2}}^{N}(\Omega_2)
 I_{\xi_{1}}^{\ast}(-\Omega_{1})
 I_{\xi_{4}}(\Omega_{3})
\Gamma_{\xi_{4}\xi_{1}\xi_{3}\xi_{2}}(\Omega_{3} + \mathcs{e}_{\xi_{1}}+i\gamma_{\xi_{1}})
\mathcal{I}_{\xi_{3}\xi_{2}}(\Omega_{3} + \mathcs{e}_{\xi_{1}}+i\gamma_{\xi_{1}}),
\label{eq:SignalIntK1_fin}
\end{eqnarray}
\begin{eqnarray}
 &  & \mathbb{S}_{\nu_{4}\nu_{3}\nu_{2}\nu_{1}}^{\mathbf{k}_{II}}
(\Omega_3,\Omega_2,\Omega_1)=
2i\sum_{\xi_{4}...\xi_{1}}
\langle
     \mathbs{d}_{\xi_{4}}^{\nu_{4}}(\mathbf{k}_{2}-\mathbf{k}_{3}-\mathbf{k}_{1})
	 \mathbs{d}_{\xi_{3}}^{\nu_{3}\ast}(-\mathbf{k}_{3})
	 \mathbs{d}_{\xi_{2}}^{\nu_{2}}(-\mathbf{k}_{2})
	 \mathbs{d}_{\xi_{1}}^{\nu_{1}\ast}(-\mathbf{k}_{1})
\rangle\nonumber \\
 &  & \quad\quad\times
I_{\xi_{2}\xi_{1}}^{N}(\Omega_2)
I_{\xi_{1}}(\Omega_{1})
I_{\xi_{4}}(\Omega_{3})
 \Gamma_{\xi_{4}\xi_{2},\xi_{3}\xi_{1}}(\Omega_3+\mathcs{e}_{\xi_{2}}+i\gamma_{\xi_{2}})
 \mathcal{I}_{\xi_{3}\xi_{1}}(\Omega_3+\mathcs{e}_{\xi_{2}}+i\gamma_{\xi_{2}}),
 \label{eq:SignalIntK2_fin}
\end{eqnarray}
\begin{eqnarray}
 &  & \mathbb{S}_{\nu_{4}\nu_{3}\nu_{2}\nu_{1}}^{\mathbf{k}_{III}}
 (\Omega_3,\Omega_2,\Omega_1)=
 2i\sum_{\xi_{4}...\xi_{1}}
 \langle
     \mathbs{d}_{\xi_{4}}^{\nu_{4}}(\mathbf{k}_{3}-\mathbf{k}_{2}-\mathbf{k}_{1})
	 \mathbs{d}_{\xi_{3}}^{\nu_{3}}(-\mathbf{k}_{3})
	 \mathbs{d}_{\xi_{2}}^{\nu_{2}\ast}(-\mathbf{k}_{2})
	 \mathbs{d}_{\xi_{1}}^{\nu_{1}\ast}(-\mathbf{k}_{1})
\rangle\nonumber \\
 &  & \quad\quad\times
 I_{\xi_{1}}(\Omega_1)
 I_{\xi_{4}}(\Omega_{3})
 I_{\xi_{3}}^{\ast}(\Omega_2-\Omega_{3}) [
 \Gamma_{\xi_{4}\xi_{3},\xi_{2}\xi_{1}}(\Omega_2)
 \mathcal{I}_{\xi_{2}\xi_{1}}(\Omega_2) \nonumber \\
 &  & \qquad \qquad - \Gamma_{\xi_{4}\xi_{3},\xi_{2}\xi_{1}}(\Omega_3+\mathcs{e}_{\xi_3}+i\gamma_{\xi_3})
 \mathcal{I}_{\xi_{2}\xi_{1}}(\Omega_3+\mathcs{e}_{\xi_3}+i\gamma_{\xi_3})],\label{eq:SignalIntK3_fin}
\end{eqnarray}
where we have defined the exciton transition dipole for state $\xi$
\begin{equation}
\mathbs{d}_{\xi}^{\nu}(\mathbf{k})=\sum_{m}e^{i\mathbf{k}\mathbf{r}_{m}}\mathbs{\mu}_{m}^{\nu}\psi_{\xi m}\label{eq:ExcitonMiuK}
\end{equation}
and 
\begin{equation}
I_{\xi_{2}\xi_{1}}^{N}(\Omega) = \frac{i}{\Omega+\mathcs{e}_{\xi_2}-\mathcs{e}_{\xi_1}+i(\gamma_{\xi_2}+\gamma_{\xi_1})}.
\end{equation}

The application of the Green's functions expressions for periodic
infinite systems is straightforward \cite{Abramavicius2005jcp-chiral}: the summations over one-exciton
eigenstates are replaced by summations over different Davydov
exciton bands at zero momentum. The scattering matrix of 
infinite systems which  involves all possible momenta of different Davydov
bands is given in appendix \ref{sec:The-exciton-infinite}.

\section{Application to the amide I band of peptides}

\label{sec:Application-to-peptides}

We have applied the present theory to the amide I vibrations of two
ideal structural motifs of polypeptides: $\alpha$ helices (one
dimensional) with 18 amide residues in the unit cell and
antiparallel $\beta$ sheets (two dimensional) with 4 residues per
unit cell. The structural and coupling parameters were
reported earlier
\cite{AbramaviciusMukamel2004,Abramavicius2005jcp-chiral}. The
anharmonicity $\Delta_{mn}$ is local ($\Delta$ = -16 cm\sups{-1} for
$m=n$ and zero otherwise) and the same dephasing rate, $\gamma_{\xi}$
= 3 cm\sups{-1} was assumed for all excitons. We used 100 cells in each dimension to
calculate the scattering matrix with periodic boundary conditions
(eq. (\ref{eq:ScatMatrixFinal}) ).

The linear absorption of both motifs presented in figure
\ref{enu:SignalLA} shows two absorption peaks for the $\alpha$ helix and
the $\beta$ sheet. For the $\alpha$ helix the longitudinal
transition -- along the helix axis gives a peak at 1642 cm\sups{-1},
and  transverse -- in the plane perpendicular to the helical axis
gives a weaker peak at  1661 cm\sups{-1} \cite{AbramaviciusMukamel2004}.  For the $\beta$ sheet
the main peak at 1632 cm\sups{-1} is horizontal (parallel to the sheet surface).
The other weak peak at 1700 cm\sups{-1} consists of two transitions \cite{AbramaviciusMukamel2004}:
vertical (perpendicular to the sheet surface transition) at
1707 cm\sups{-1} and horizontal at 1699 cm\sups{-1}.

Signals were calculated for a collinear field
configuration where, $\mathbf{k}_{1}$, $\mathbf{k}_{2}$ and
$\mathbf{k}_{3}$, propagate along $z$. We further assume all fields
to have the same carrier frequency so that
$\left|\mathbf{k}_{j}\right|=\omega_{s}/c$ for all $\mathbf{k}_{j}$
($j=1,2,3$) where $c$ is the speed of light. The absolute magnitudes of 
the $\mathbb{S}_{\nu_{4}\nu_{3}\nu_{2}\nu_{1}}^{\mathbf{k}_{I}}(\Omega_{3},T_2=0,\Omega_{1})$
and
$\mathbb{S}_{\nu_{4}\nu_{3}\nu_{2}\nu_{1}}^{\mathbf{k}_{III}}(\Omega_{3},\Omega_{2},T_1=0)$
signals were computed using eqs. (\ref{eq:SignalIntK1}) and
(\ref{eq:SignalIntK3}). These are one-sided Fourier transforms of the time
domain $\mathbf{k}_{I}$ and $\mathbf{k}_{III}$ signals.

The $\mathbb{S}_{\nu_{4}\nu_{3}\nu_{2}\nu_{1}}^{\mathbf{k}_{I}}$ signals for
the $\alpha$ helix are shown in the left column in figure
\ref{enu:SignalK1}. The $xxxx$ component is finite in the dipole
approximation and shows one major peak associated with the
longitudinal transition \cite{AbramaviciusMukamel2004} and a much weaker
20 cm\sups{-1} blue shifted transverse transition.
The crosspeaks between these two peaks are symmetric with respect
to the diagonal and have roughly the same amplitudes.
This signal resembles our previous calculation of large (90 residue)
helices \cite{AbramaviciusMukamel2004}, which could be treated as
infinite helices with small edge defects.

This $xxxx$ pattern is changed in the
chirally-sensitive $xxxy$ component. The diagonal peaks are
suppressed.
This can be explained by noting that the distance
between sites enters the signal amplitude: diagonal peaks have no
such distance factor since they originate from interactions
with the same mode.  The crosspeaks are induced by interactions of
different modes and depend on their distance. Thus, in
the chiral signal the crosspeaks are amplified compared to the
diagonal peaks. Additionally, the crosspeak pattern is asymmetric
since one of the crosspeaks (below the diagonal) dominates. The
other, above the diagonal, crosspeak is much weaker. The remaining two
chiral components $xxyx$ and $xyxx$  also suppress
the diagonal peaks, while show two crosspeaks

The antiparallel $\beta$ sheet signals displayed in the right column
in figure \ref{enu:SignalK1} show a similar pattern, however only
one horizontal diagonal peak (as shown in
\cite{AbramaviciusMukamel2004}) is visible in $xxxx$ with very weak
crosspeaks shifted to much higher frequencies compared to the
diagonal. Since there are four sites per unit cell, we generally
expect four optical transitions (Davydov components). This could
result in four diagonal peaks and six crosspeaks in both sides of
the diagonal if all transitions are correlated. However, $xxxx$ is
dominated by a single strong horizontal transition. Similar to the
helix, this peak structure changes in the chiral component, $xxxy$,
which shows much stronger crosspeaks than the diagonal peaks. The
strongest crosspeak is well separated from the strongest $xxxx$
peak. Like the helix, the $xxyx$ and $xyxx$ configurations are both
similar and show crosspeaks above and below the diagonal.

We next turn to the $\mathbf{k}_{III}$ technique.
$\mathbb{S}_{\nu_{4}\nu_{3}\nu_{2}\nu_{1}}^{\mathbf{k}_{III}}$ shows
two-exciton resonances along $\omega_{2}$ (double quantum
coherence); the resonances along $\omega_{3}$ originate both from
one- to two-exciton resonances ($e$ to $f$ in figure
\ref{enu:Scheme}) and from one-exciton to ground state resonances
($0$ to $e$). The $xxxx$ component for the $\alpha$ helix presented
in figure \ref{enu:SignalK2} shows one major peak and a weaker peak
originating from the same two-exciton resonance. The main peak of
the chiral components $xxxy$, $xxyx$, $xyxx$ is shifted compared to
$xxxx$, indicating that different two-exciton resonances make the
strongest contribution to the signal. The $xxxy$ and $xxyx$ signals
are very similar and show two strong peaks unlike $xyxx$ which only shows one major peak. The $xxxx$ component of the $\beta$
sheet displayed in figure \ref{enu:SignalK2} has a similar
structure to the $\alpha$ helix. The chiral $xxxy$ component shows two peaks at higher two-exciton resonance comparing to $xxxx$. $xxyx$, is very similar. $xyxx$ shows one major peak at $\omega_{3}$ = ~1700 cm\sups{-1}.

\section{Discussion}

\label{sec:Discussion}

In the electric dipole approximation, the linear absorption of
isotropic systems is related to the diagonal tensor elements of the
linear susceptibility
\cite{wagersreitermukamel,Abramavicius2005jcp-chiral}. Nonlinear
signals calculated within this approximation then provide a limited
window onto the optical responses of isotropic ensembles through
three independent chirally-non-sensitive tensor components of the
third order response function: $xxyy$, $xyyx$, $xyxy$. The signal
propagation direction is determined by phase-matching. Going beyond
the dipole approximation, we found three additional components for a
collinear laser configuration: $xxxy$, $xxyx$, $xyxx$. Noncollinear
configurations (which can also satisfy phase matching) lead to six
nonzero elements (the three additional components are: $(z)xyzz$,
$(z)xzzy$ and $(z)xzyz$, where the first index $(z)$ denotes the
wavevector, and the other four are polarization
components). These are chirally--sensitive and show a 
different pattern in the correlation plots, which reflects the
polarization properties of optical transitions. For example, from
figures \ref{enu:SignalK1} and \ref{enu:SignalK2} we see that the
strongest peaks originate from correlations between perpendicular
transitions: longitudinal - transverse in the $\alpha$ helix and
horizontal - vertical in the $\beta$ sheet (see figure
\ref{enu:Scheme}).

The differences between the chiral components can be deduced by
following the interaction and evolution sequences: in
$\mathbb{S}^{\mathbf{k}_{I}}$, $\Omega_{1}$ corresponds to the free 
evolution after the first interaction. Subsequently two simultaneous
interactions take place. Therefore $xxyx$ and $xyxx$ reflect a
similar excitation pattern: there is one interaction first, followed
by two interactions with the perpendicular polarization.  $xxxy$ is
qualitatively different: the first interaction is followed by two
interactions with parallel polarizations. This is why, $xxxy$ and
$xxyx$ show a different peak pattern. For $\mathbb{S}^{\mathbf{k}_{III}}$ there are two simultaneous
interactions first, followed by one interaction. Thus, $xxxy$ and
$xxyx$ give very similar signals since in both cases the first two
interactions are with perpendicular fields. The $xyxx$ component is
qualitatively different since the first two interactions are with
parallel fields.

The signal further carries information regarding the redistribution of
excitonic amplitudes between states with different polarization
properties. For instance, $\mathbb{S}_{xyxx}^{\mathbf{k}_{III}}$ originates from exciton amplitude transfer
between a state with $x$ polarization to a state with $y$
polarization during $T_2$. $\mathbb{S}_{xxxy}^{\mathbf{k}_{III}}$ is qualitatively different since the
evolution during $T_2$ corresponds to $x$ polarized exciton
amplitude transfer to another $x$ exciton. $\mathbb{S}_{xxxy}^{\mathbf{k}_{I}}$ again originates from exciton amplitude
transfer during $T_1$ between states with perpendicular
($y\rightarrow x$) polarizations; $xyxx$ does not require such
exciton evolution.

In the present work we only included homogeneous line
broadening caused by fast frequency fluctuations. Slow 
fluctuations result in inhomogeneous broadening which should affect
the 2D lineshapes \cite{HammHochstrasser1998,DijkstraKnoester2005}. 
The ideal peak patterns predicted for
periodic structures can be used for structure decomposition
and parameter determination of real structures. Dynamical
information on exciton evolution can be obtained by varying $T_2$
in $\mathbb{S}^{\mathbf{k}_{I}}(\Omega_{3},T_2,\Omega_{1})$. Population
transport may be incorporated using the theory developed in \cite{Chernyak_physrep1995,MukamelAbramavicius2004}, but this goes beyond the scope of this
paper.

We next compare the exciton scattering mechanism offered by the NEE
with the more conventional picture of transitions among eigenstates,
as described by double-sided Feynman diagrams. The relation between
the two for $\mathbf{k}_{I}$, $\mathbf{k}_{II}$ and
$\mathbf{k}_{III}$ is shown in figure \ref{enu:Comparison}. For
$\mathbf{k}_{I}$ the exciton coherence during $T_1$ in both
pictures is represented by the same Green's function. However,
during $T_3$ the Feynman diagrams show two independent coherence
evolutions, only one of them involves the two-exciton states. The
scattering representation is more complex: during $T_3$ the
evolution consists of scattering + two free evolutions. This is
natural in the molecular basis set where the exciton pathways can be
followed in real space: the short range exciton scattering is then
followed by free evolution as the excitons separate. The role of the 
distance parameter is much less obvious in the eigenstate picture. $\mathbf{k}_{II}$ is
similar, except the density matrix evolution during the first and
the last intervals has the same frequency sign. In
$\mathbf{k}_{III}$ two excitons are created by the initial
excitation. Thus the scattering can occur at any time. In the sum over states representation this is given as independent evolutions of
two-exciton coherences during $T_1$. Both representations of the
response are equivalent as long as pure dephasing is neglected. The
NEE and the corresponding response become more complicated by pure
dephasing which induces incoherent population transport and requires additional dynamic variables \cite{Chernyak_physrep1995,CherniakMukamel96,MukamelAbramavicius2004}.

The response function in eq. (\ref{eq:ThirdResponse}) is not required to be time ordered, thus, we can define it to be symmetric with respect to permutation of different time arguments. This choice is useful for overlapping optical pulses and frequency domain response, where time ordering of interactions is not enforced. Frequency domain signals can be expressed using the third order susceptibility given in terms of the exciton scattering matrix as shown in appendix
\ref{sec:susceptibilities}. The susceptibility is then directly related to non time ordered response function. Expressions for $\mathbf{k}_I$, $\mathbf{k}_{II}$ and $\mathbf{k}_{III}$ can be derived using the RWA.

Our theory may be applied to other periodic (infinite) as well as to
non periodic (finite) systems. Periodicity and cyclic boundary
conditions result in exciton band structure where only zero momenta
exciton states are active. This applies as long as the optical
wavevector is small compared with the exciton momentum.
Exciton-exciton interactions are the source of anharmonicities and
are described using quasi-particle scattering. This description
may also be applied to semiconductors and quantum superlattices
\cite{AxtMukamel1998rev,Chemla2001_semi_rev,Ogawa_semi2004,Morawetz2004}. When
the exciton coherence size becomes comparable to the optical
wavelength the theory needs to be extended to include polariton effects
\cite{KnoesterMukamel1991}.

\section{Acknowledgements}

This material is based upon work supported by the National Science
Foundation grant no. CHE-0132571 and the National Institutes of
Health grant no. 1 RO1 GM59230-10A2.  This support is gratefully
acknowledged. We also wish to thank Wei Zhuang, for most useful
discussions.

\appendix

\section{Linear absorption of excitons}

\label{sec:LASignal}

In this appendix we connect the linear absorption with the response
function. The linear response function and the linear polarization
were defined by eq. (\ref{eq:LinearResponse}). The linear absorption
can be calculated using the linear susceptibility $\chi^{(1)}$ \cite{Abramavicius2005jcp-chiral}.  The linear response, especially
for vibrations, is often measured in the time domain using
ultrashort laser pulses: Fourier transform of the time dependent 
induced polarization (free induction decay) gives the Fourier Transform
Infrared (FTIR) spectrum.

We assume a single excitation pulse with envelope $\tilde{E}_{\nu}^{(0)}(t-t_{0})$ centered at
$t_{0}$ with the carrier frequency $\omega_{0}$ and wavevector
$\mathbf{k}_{0}$:
\begin{equation}
\mathbs{E}_{\nu}(\mathbf{x})=\frac{1}{2}\delta_{\nu\nu_{0}}\tilde{E}_{\nu_{0}}^{(0)}(t-t_{0})\exp(i\mathbf{k}_{0}\mathbf{r}-i\omega_{0}(t-t_{0}))+c.c.,\label{eq:FieldLin}
\end{equation}
The linear response function of the excitonic
system is given in terms of the one-exciton Green's functions
\cite{Abramavicius2005jcp-chiral}:
\begin{equation}
S_{\nu_{2}\nu_{1}}^{(1)}(\bar{\mathbf{x}}_{2},\bar{\mathbf{x}}_{1})=i\sum_{\xi}\langle\mathbs{d}_{\xi}^{\nu_{2}}(\mathbf{k}_{2})\mathbs{d}_{\xi}^{\nu_{1}\ast}(-\mathbf{k}_{1})\rangle I_{\xi}(t_{2}-t_{1})+c.c.'\label{eq:LinRespFu}
\end{equation}
Within the dipole approximation 
rotational averaging (eq. (\ref{eq:LinRespFu})) gives
$\langle\mathbs{d}_{\xi}^{\nu_{2}}(\mathbf{k}_{2})\mathbs{d}_{\xi}^{\nu_{1}\ast}(-\mathbf{k}_{1})\rangle\approx\frac{1}{3}\delta_{\nu_{2}\nu_{1}}|\mathbs{d}_{\xi}|^{2}$
\cite{Abramavicius2005jcp-chiral}. Substituting eqs.
(\ref{eq:FieldLin}) and (\ref{eq:LinRespFu}) into eq.
(\ref{eq:LinearResponse}) and using assumptions for the field given in
section \ref{sec:NonLinearAbsorption} we can integrate $\int
d\mathbf{x}_{1}$ which leads to:
\begin{equation}
\mathbs{P}_{\nu_{0}}(\mathbf{k}_{2}t_{2})=\frac{i}{6}E_{\nu_{0}}^{(0)}\sum_{\xi}|\mathbs{d}_{\xi}|^{2}[\delta(\mathbf{k}_{2}+\mathbf{k}_{0})I_{\xi}(t_{2}-t_{1})-\delta(\mathbf{k}_{2}-\mathbf{k}_{0})I_{\xi}^{*}(t_{2}-t_{1})],\label{eq:Pol1Fin}
\end{equation}
where $\delta(\mathbf{k})$ accounts for the translational invariance of an 
isotropic system. This expression shows that
$\mathbs{P}_{\nu_{0}}(\mathbf{k}_{2}t_{2})$ only depends on the
delay $T_1=t_2-t_1$ and can, thus, be denoted
$\mathbs{P}_{\nu_{0}}(\mathbf{k}_{2}t_{2}) \equiv
\mathbs{P}_{\nu_{0}}(\mathbf{k}_{2},T_1)$. The first term in eq.
(\ref{eq:Pol1Fin}) in real space is proportional to
$e^{i\mathbf{k}_{0}\mathbf{r}}$ and, thus, describes forward
propagation of the induced polarization.

The absorption spectrum is defined as the imaginary part of
Fourier transform of the linear polarization with respect to
$T_1$: $\sigma_A(\omega) = Im \int_0^{\infty} dT_1 \exp(i\omega T_1)  \mathbs{P}_{\nu_{0}}(\mathbf{k}_{2},T_1) $,
while the real part of this integral describes dispersive lineshapes. By eliminating the
prefactor
$(1/6)E_{\nu_{0}}^{(0)}\delta(\mathbf{k}_{2}+\mathbf{k}_{0})$ and
keeping only terms resonant to the positive frequency we obtain:
\begin{equation}
\sigma_{A}(\omega)=\sum_{\xi}\frac{\gamma_{\xi}|\mathbs{d}_{\xi}|^{2}}{(\omega-\mathcs{e}_{\xi})^{2}+\gamma_{\xi}^{2}}.\label{eq:LinAbs}
\end{equation}

The CD spectrum must be calculated by going beyond the dipole approximation
\cite{Abramavicius2005jcp-chiral}.

\section{The response of isotropic ensembles}

\label{sec:RotationalAveraging}

Optical fields, wavevectors, and space coordinates are defined in
the lab frame, while the transition dipoles and their position vectors
are given in the molecular frame. The rotational averaging in
eqs. (\ref{eq:TimeOrderedResponse1Fin}) and (\ref{eq:TimeOrderedResponse3Fin}), $\langle...\rangle$,
needs to be performed over the relative orientation of the two frames
in order to calculate the response functions for isotropic (randomly
oriented) ensembles of molecules \cite{andrewsthirumanchandran77jcpROT,Barron-CD-book-1982}:
\begin{eqnarray}
 &  & \langle\mathbs{d}_{\xi_{4}}^{\nu_{4}}(\mathbf{k}_{4})\mathbs{d}_{\xi_{3}}^{\nu_{3}}(\mathbf{k}_{3})\mathbs{d}_{\xi_{2}}^{\nu_{2}\ast}(-\mathbf{k}_{2})\mathbs{d}_{\xi_{1}}^{\nu_{1}\ast}(-\mathbf{k}_{1})\rangle\nonumber \\
 &  & \quad\quad=\sum_{n_{4}n_{3}n_{2}n_{1}}\psi_{\xi_{4}n_{4}}\psi_{\xi_{3}n_{3}}\psi_{\xi_{2}n_{2}}^{\ast}\psi_{\xi_{1}n_{1}}^{\ast}\nonumber \\
 &  & \quad\quad\times\langle e^{i\mathbf{k}_{4}\mathbf{r}_{n_{4}}+i\mathbf{k}_{3}\mathbf{r}_{n_{3}}+i\mathbf{k}_{2}\mathbf{r}_{n_{2}}+i\mathbf{k}_{1}\mathbf{r}_{n_{1}}}\mathbs{\mu}_{n_{4}}^{\nu_{4}}\mathbs{\mu}_{n_{3}}^{\nu_{3}}\mathbs{\mu}_{n_{2}}^{\nu_{2}}\mathbs{\mu}_{n_{1}}^{\nu_{1}}\rangle,\label{eq:AveragingExcitonDipoles3}
 \end{eqnarray}
 In the phase-matching directions the microscopic response functions
only depend on the relative positions of molecules and are
independent on the origin of the coordinate system. Since the
coordinates $\mathbf{r}_{m}$ vary only within one molecule, for
molecules smaller than the wavelength of light we have
$\mathbf{k}\mathbf{r}_{m} \ll 1$, and the exponential function in
the transition dipole of eq. (\ref{eq:ExcitonMiuK}) can be expanded
to first order. This leads to
\begin{eqnarray}
 &  & \langle e^{i\mathbf{k}_{4}\mathbf{r}_{n_{4}}+i\mathbf{k}_{3}\mathbf{r}_{n_{3}}+i\mathbf{k}_{2}\mathbf{r}_{n_{2}}+i\mathbf{k}_{1}\mathbf{r}_{n_{1}}}\mathbs{\mu}_{n_{4}}^{\nu_{4}}\mathbs{\mu}_{n_{3}}^{\nu_{3}}\mathbs{\mu}_{n_{2}}^{\nu_{2}}\mathbs{\mu}_{n_{1}}^{\nu_{1}}\rangle\nonumber \\
 &  & \qquad\approx\langle\mathbs{\mu}_{n_{4}}^{\nu_{4}}\mathbs{\mu}_{n_{3}}^{\nu_{3}}\mathbs{\mu}_{n_{2}}^{\nu_{2}}\mathbs{\mu}_{n_{1}}^{\nu_{1}}\rangle\nonumber \\
 &  & \qquad+i\sum_{\kappa}\mathbf{k}_{4}^{\kappa}\langle\mathbf{r}_{n_{4}}^{\kappa}\mathbs{\mu}_{n_{4}}^{\nu_{4}}\mathbs{\mu}_{n_{3}}^{\nu_{3}}\mathbs{\mu}_{n_{2}}^{\nu_{2}}\mathbs{\mu}_{n_{1}}^{\nu_{1}}\rangle\nonumber \\
 &  & \qquad+i\sum_{\kappa}\mathbf{k}_{3}^{\kappa}\langle\mathbf{r}_{n_{3}}^{\kappa}\mathbs{\mu}_{n_{4}}^{\nu_{4}}\mathbs{\mu}_{n_{3}}^{\nu_{3}}\mathbs{\mu}_{n_{2}}^{\nu_{2}}\mathbs{\mu}_{n_{1}}^{\nu_{1}}\rangle\nonumber \\
 &  & \qquad+i\sum_{\kappa}\mathbf{k}_{2}^{\kappa}\langle\mathbf{r}_{n_{2}}^{\kappa}\mathbs{\mu}_{n_{4}}^{\nu_{4}}\mathbs{\mu}_{n_{3}}^{\nu_{3}}\mathbs{\mu}_{n_{2}}^{\nu_{2}}\mathbs{\mu}_{n_{1}}^{\nu_{1}}\rangle\nonumber \\
 &  & \qquad+i\sum_{\kappa}\mathbf{k}_{1}^{\kappa}\langle\mathbf{r}_{n_{1}}^{\kappa}\mathbs{\mu}_{n_{4}}^{\nu_{4}}\mathbs{\mu}_{n_{3}}^{\nu_{3}}\mathbs{\mu}_{n_{2}}^{\nu_{2}}\mathbs{\mu}_{n_{1}}^{\nu_{1}}\rangle.\label{eq:AverageMolecExpanded3}
 \end{eqnarray}
 In the exciton basis, eqs. (\ref{eq:AveragingExcitonDipoles3}) and (\ref{eq:AverageMolecExpanded3}) give
\begin{eqnarray}
 &  & \langle\mathbs{d}_{\xi_{4}}^{\nu_{4}}(\mathbf{k}_{4})\mathbs{d}_{\xi_{3}}^{\nu_{3}}(\mathbf{k}_{3})\mathbs{d}_{\xi_{2}}^{\nu_{2}\ast}(-\mathbf{k}_{2})\mathbs{d}_{\xi_{1}}^{\nu_{1}\ast}(-\mathbf{k}_{1})\rangle\nonumber \\
 &  & \qquad=\langle\mathbs{d}_{\xi_{4}}^{\nu_{4}}\mathbs{d}_{\xi_{3}}^{\nu_{3}}\mathbs{d}_{\xi_{2}}^{\nu_{2}}\mathbs{d}_{\xi_{1}}^{\nu_{1}}\rangle\nonumber \\
 &  & \qquad+i\sum_{\kappa}\mathbf{k}_{4}^{\kappa}\langle\bar{\mathbs{d}}_{\xi_{4}}^{\kappa,\nu_{4}}\mathbs{d}_{\xi_{3}}^{\nu_{3}}\mathbs{d}_{\xi_{2}}^{\nu_{2}}\mathbs{d}_{\xi_{1}}^{\nu_{1}}\rangle\nonumber \\
 &  & \qquad+i\sum_{\kappa}\mathbf{k}_{3}^{\kappa}\langle\bar{\mathbs{d}}_{\xi_{3}}^{\kappa,\nu_{3}}\mathbs{d}_{\xi_{4}}^{\nu_{4}}\mathbs{d}_{\xi_{2}}^{\nu_{2}}\mathbs{d}_{\xi_{1}}^{\nu_{1}}\rangle\nonumber \\
 &  & \qquad+i\sum_{\kappa}\mathbf{k}_{2}^{\kappa}\langle\bar{\mathbs{d}}_{\xi_{2}}^{\kappa,\nu_{2}}\mathbs{d}_{\xi_{3}}^{\nu_{3}}\mathbs{d}_{\xi_{4}}^{\nu_{4}}\mathbs{d}_{\xi_{1}}^{\nu_{1}}\rangle\nonumber \\
 &  & \qquad+i\sum_{\kappa}\mathbf{k}_{1}^{\kappa}\langle\bar{\mathbs{d}}_{\xi_{1}}^{\kappa,\nu_{1}}\mathbs{d}_{\xi_{3}}^{\nu_{3}}\mathbs{d}_{\xi_{2}}^{\nu_{2}}\mathbs{d}_{\xi_{4}}^{\nu_{4}}\rangle,\label{eq:rotAv3ExcExp}
 \end{eqnarray}
where we have used the fact that $\psi_{\xi,m}$ are real and have
defined the transition dipole vector for zero momentum exciton state
\begin{equation}
\mathbs{d}_{\xi}^{\nu}\equiv\mathbs{d}_{\xi}^{\nu}(\mathbf{k}=0)=\sum_{m}\mathbs{\mu}_{m}^{\nu}\psi_{\xi,m},\label{eq:ExcitonDipole0}
\end{equation}
 and the tensor
 \begin{equation}
\bar{\mathbs{d}}_{\xi}^{\kappa,\nu}=\sum_{m}\mathbf{r}_{m}^{\kappa}\mathbs{\mu}_{m}^{\nu}\psi_{\xi,m}.\label{eq:ExcitonTensor0}
\end{equation}

Eq. (\ref{eq:rotAv3ExcExp}) requires fourth and fifth rank
rotational averagings. The first term in this equation corresponds
to the dipole approximation. The remaining terms which contain the
wavevector and a coordinate, represent a first order correction to
that approximation. These terms do not depend on the origin of
the molecular frame provided
$\mathbf{k}_{4}+\mathbf{k}_{3}+\mathbf{k}_{2}+\mathbf{k}_{1}=0$ (i. e. the
signal wavevector $\mathbf{k}_{S}=-\mathbf{k}_{4}$), which is the
phase matching condition.

Rotational averaging can be performed using the transformation
between the lab and molecular frames
\cite{Barron-CD-book-1982}:
\begin{equation}
\langle\mathbf{a}_{s}^{\nu_{s}}...\mathbf{a}_{1}^{\nu_{1}}\rangle\equiv\langle(\mathbf{e}^{\nu_{s}}\cdot\mathbf{a}_{s})...(\mathbf{e}^{\nu_{1}}\cdot\mathbf{a}_{1})\rangle=\sum_{\alpha_{s}...\alpha_{1}}\mathbf{T}_{\nu_{s}...\nu_{1},\alpha_{s}...\alpha_{1}}^{(s)}\mathbf{a}_{s}^{\alpha_{s}}...\mathbf{a}_{1}^{\alpha_{1}},\label{eq:SimpleAveraging}
\end{equation}
 where $\mathbf{T}_{\nu_{s}...\nu_{1},\alpha_{s}...\alpha_{1}}^{(s)}=\langle l_{\nu_{s}\alpha_{s}}...l_{\nu_{1}\alpha_{1}}\rangle$
is the average of transformation tensor where $l_{\nu\alpha}$ is the
cosine of the angle between laboratory frame axis $\nu=x,y,z$ and
molecular frame axis $\alpha=x,y,z$. The necessary averages of ranks
four and five transformation tensors, which are universal quantities
independent of system geometry are given in table
\ref{tab:Isotropic-rotational-average}
\cite{andrewsthirumanchandran77jcpROT}.

Using table \ref{tab:Isotropic-rotational-average}, we obtain for
the rotational averages of the transition dipoles:
\begin{equation}
\langle\mathbs{d}_{\xi_{4}}^{\nu_{4}}\mathbs{d}_{\xi_{3}}^{\nu_{3}}\mathbs{d}_{\xi_{2}}^{\nu_{2}}\mathbs{d}_{\xi_{1}}^{\nu_{1}}\rangle=\sum_{\alpha_{4}\alpha_{3}\alpha_{2}\alpha_{1}}\mathbf{T}_{\nu_{4}\nu_{3}\nu_{2}\nu_{1},\alpha_{4}\alpha_{3}\alpha_{2}\alpha_{1}}^{(4)}\mathbs{d}_{\xi_{4}}^{\alpha_{4}}\mathbs{d}_{\xi_{3}}^{\alpha_{3}}\mathbs{d}_{\xi_{2}}^{\alpha_{2}}\mathbs{d}_{\xi_{1}}^{\alpha_{1}},\label{eq:RotAv4}
\end{equation}
\begin{equation}
\langle\bar{\mathbs{d}}_{\xi_{4}}^{\kappa,\nu_{4}}\mathbs{d}_{\xi_{3}}^{\nu_{3}}\mathbs{d}_{\xi_{2}}^{\nu_{2}}\mathbs{d}_{\xi_{1}}^{\nu_{1}}\rangle=\sum_{\alpha_{5}...\alpha_{1}}\mathbf{T}_{\kappa\nu_{4}\nu_{3}\nu_{2}\nu_{1},\alpha_{5}\alpha_{4}\alpha_{3}\alpha_{2}\alpha_{1}}^{(5)}\bar{\mathbs{d}}_{\xi_{4}}^{\alpha_{5},\alpha_{4}}\mathbs{d}_{\xi_{3}}^{\alpha_{3}}\mathbs{d}_{\xi_{2}}^{\alpha_{2}}\mathbs{d}_{\xi_{1}}^{\alpha_{1}}.\label{eq:RotAv5}
\end{equation}
 The remaining averages can be simply obtained by permutation of indices.

\section{Time-domain FWM signals}

\label{sec:TDSignals}

The response functions (eqs. (\ref{eq:TOrd1}) - (\ref{eq:TOrd3}) ) can be readily 
calculated  using the single exciton basis (eq. (\ref{eq:1ExEigenEquation}) ) where the number of terms
in the response function is considerably reduced. We define:
\begin{eqnarray}
\mathcal{J}_{\xi_{4}\xi_{3}\xi_{2}\xi_{1}}(\tau_{3},\tau_{2},\tau_{1}) & = & \int_{0}^{\tau_{2}}d\tau_{s}''\int_{0}^{\tau_{s}''}d\tau_{s}'\Gamma_{\xi_{4}\xi_{3},\xi_{2}\xi_{1}}(\tau_{s}''-\tau_{s}')\nonumber \\
 & \times & I_{\xi_{4}}(\tau_{s}')I_{\xi_{3}}^{\ast}(\tau_{3}-\tau_{s}')I_{\xi_{2}}(\tau_{2}-\tau_{s}'')I_{\xi_{1}}(\tau_{1}-\tau_{s}'')\label{eq:FunctionJ}
\end{eqnarray}
with the eigenstate basis scattering matrix:
\begin{equation}
\Gamma_{\xi_{4}\xi_{3}\xi_{2}\xi_{1}}(\tau)=\sum_{m_{4}...m_{1}}\psi_{\xi_{4}m_{4}}^{\ast}\psi_{\xi_{3}m_{3}}^{\ast}\Gamma_{m_{4}m_{3}m_{2}m_{1}}(\tau)\psi_{\xi_{2}m_{2}}\psi_{\xi_{1}m_{1}}.\label{eqApp:ScatMatricExcBasis}
\end{equation}
By transforming the coordinates to momentum space, we obtain from  eqs. (\ref{eq:TOrd1}) - (\ref{eq:TOrd3}):
\begin{eqnarray}
 &  & S_{\nu_{4}...\nu_{1}}^{(I)}(\bar{\mathbf{x}}_{4},...,\bar{\mathbf{x}}_{4})\nonumber \\
 &  & \quad\quad=2i\sum_{\xi_{4}...\xi_{1}}\langle\mathbs{d}_{\xi_{4}}^{\nu_{4}}(\mathbf{k}_{4})\mathbs{d}_{\xi_{3}}^{\nu_{3}\ast}(-\mathbf{k}_{3})\mathbs{d}_{\xi_{2}}^{\nu_{2}\ast}(-\mathbf{k}_{2})\mathbs{d}_{\xi_{1}}^{\nu_{1}}(\mathbf{k}_{1})\rangle\mathcal{J}_{\xi_{4}\xi_{1}\xi_{3}\xi_{2}}(t_{41},t_{43},t_{42}),\label{eq:TimeOrderedResponse1Fin}
 \end{eqnarray}
\begin{eqnarray}
 &  & S_{\nu_{4}...\nu_{1}}^{(II)}(\bar{\mathbf{x}}_{4},...,\bar{\mathbf{x}}_{4})\nonumber \\
 &  & \quad\quad=2i\sum_{\xi_{4}...\xi_{1}}\langle\mathbs{d}_{\xi_{4}}^{\nu_{4}}(\mathbf{k}_{4})\mathbs{d}_{\xi_{3}}^{\nu_{3}\ast}(-\mathbf{k}_{3})\mathbs{d}_{\xi_{2}}^{\nu_{2}}(\mathbf{k}_{2})\mathbs{d}_{\xi_{1}}^{\nu_{1}\ast}(-\mathbf{k}_{1})\rangle\mathcal{J}_{\xi_{4}\xi_{2}\xi_{3}\xi_{1}}(t_{42},t_{43},t_{41}),\label{eq:TimeOrderedResponse2Fin}
 \end{eqnarray}
\begin{eqnarray}
 &  & S_{\nu_{4}...\nu_{1}}^{(III)}(\bar{\mathbf{x}}_{4},...,\bar{\mathbf{x}}_{4})\nonumber \\
 &  & \quad\quad=2i\sum_{\xi_{4}...\xi_{1}}\langle\mathbs{d}_{\xi_{4}}^{\nu_{4}}(\mathbf{k}_{4})\mathbs{d}_{\xi_{3}}^{\nu_{3}}(\mathbf{k}_{3})\mathbs{d}_{\xi_{2}}^{\nu_{2}\ast}(-\mathbf{k}_{2})\mathbs{d}_{\xi_{1}}^{\nu_{1}\ast}(-\mathbf{k}_{1})\rangle\mathcal{J}_{\xi_{4}\xi_{3}\xi_{2}\xi_{1}}(t_{43},t_{42},t_{41}).\label{eq:TimeOrderedResponse3Fin}
 \end{eqnarray}
The wavevector-dependence enters through the transition dipoles. These expressions
need to be rotationally averaged for isotropic ensembles, as
described in appendix \ref{sec:RotationalAveraging}.

Equations
(\ref{eq:TimeOrderedResponse1Fin})-(\ref{eq:TimeOrderedResponse3Fin}) can be
conveniently expressed in terms of the frequency domain scattering
matrix (eqs. (\ref{eq:ScattMatrixW}) and (\ref{eq:ScatRelationTW})
).  These expressions then involve triple integrals (two with
respect to time and one with respect to the scattering matrix
frequency). However, the integration limits in eq.
(\ref{eq:FunctionJ}) are controlled by multiple $\theta(t)$
functions coming from the Green's functions. Taking these into
account and using the frequency domain scattering matrix (eq.
(\ref{eq:ScattMatrixW})), the integrals over the exponential
functions can be considerably simplified. The integration limits
$(-\infty,\tau_{2})$ for $\tau_{s}''$ and $(0,\tau_{3})$ for
$\tau_{s}'$ always hold for $S^{I}$, $S^{II}$ and $S^{III}$ in eqs.
(\ref{eq:TimeOrderedResponse1Fin})-(\ref{eq:TimeOrderedResponse3Fin}).
However, for $S^{I}$ (eq. (\ref{eq:TimeOrderedResponse1Fin}) ) and
$S^{II}$ (eq. (\ref{eq:TimeOrderedResponse2Fin}) ) (but not for eq.
(\ref{eq:TimeOrderedResponse3Fin}) ) other integration limits
$(0,\tau_{2})$ for $\tau_{s}'$ can also be used.  These considerations allow to
calculate time integrals:
\begin{eqnarray}
 &  & S_{\nu_{4}...\nu_{1}}^{(I)}(\mathbf{\bar{x}}_{4},...,\bar{\mathbf{x}}_{1})=2\sum_{\xi_{4}...\xi_{1}}\langle\mathbs{d}_{\xi_{4}}^{\nu_{4}}(\mathbf{k}_{4})\mathbs{d}_{\xi_{3}}^{\nu_{3}\ast}(-\mathbf{k}_{3})\mathbs{d}_{\xi_{2}}^{\nu_{2}\ast}(-\mathbf{k}_{2})\mathbs{d}_{\xi_{1}}^{\nu_{1}}(\mathbf{k}_{1})\rangle\nonumber \\
 &  & \quad\quad\times I_{\xi_{1}}^{*}(t_{21})I_{\xi_{1}}^{*}(t_{32})I_{\xi_{2}}(t_{32})\int\frac{d\omega}{2\pi}\Gamma_{\xi_{4}\xi_{1}\xi_{3}\xi_{2}}(\omega)\mathcal{I}_{\xi_{3}\xi_{2}}(\omega)\frac{I_{\xi_{4}}(t_{43})-e^{-i\omega t_{43}}I_{\xi_{1}}^{*}(t_{43})}{\omega-\mathcs{e}_{\xi_{4}}-\mathcs{e}_{\xi_{1}}+i(\gamma_{\xi_{4}}-\gamma_{\xi_{1}})}\label{eqApp:TimeOrdRespFin10}
\end{eqnarray}
\begin{eqnarray}
 &  & S_{\nu_{4}...\nu_{1}}^{(II)}(\mathbf{\bar{x}}_{4},...,\bar{\mathbf{x}}_{1})=2\sum_{\xi_{4}...\xi_{1}}\langle\mathbs{d}_{\xi_{4}}^{\nu_{4}}(\mathbf{k}_{4})\mathbs{d}_{\xi_{3}}^{\nu_{3}\ast}(-\mathbf{k}_{3})\mathbs{d}_{\xi_{2}}^{\nu_{2}}(\mathbf{k}_{2})\mathbs{d}_{\xi_{1}}^{\nu_{1}\ast}(-\mathbf{k}_{1})\rangle\nonumber \\
 &  & \quad\quad\times I_{\xi_{1}}(t_{21})I_{\xi_{2}}^{*}(t_{32})I_{\xi_{1}}(t_{32})\int\frac{d\omega}{2\pi}\Gamma_{\xi_{4}\xi_{2}\xi_{3}\xi_{1}}(\omega)\mathcal{I}_{\xi_{3}\xi_{1}}(\omega)\frac{I_{\xi_{4}}(t_{43})-e^{-i\omega t_{43}}I_{\xi_{2}}^{*}(t_{43})}{\omega-\mathcs{e}_{\xi_{4}}-\mathcs{e}_{\xi_{2}}+i(\gamma_{\xi_{4}}-\gamma_{\xi_{2}})}\label{eqApp:TimeOrdRespFin20}
 \end{eqnarray}
\begin{eqnarray}
 &  & S_{\nu_{4}...\nu_{1}}^{(III)}(\mathbf{\bar{x}}_{4},...,\bar{\mathbf{x}}_{1})=2\sum_{\xi_{4}...\xi_{1}}\langle\mathbs{d}_{\xi_{4}}^{\nu_{4}}(\mathbf{k}_{4})\mathbs{d}_{\xi_{3}}^{\nu_{3}}(\mathbf{k}_{3})\mathbs{d}_{\xi_{2}}^{\nu_{2}\ast}(-\mathbf{k}_{2})\mathbs{d}_{\xi_{1}}^{\nu_{1}\ast}(-\mathbf{k}_{1})\rangle\nonumber \\
 &  & \quad\quad\times I_{\xi_{1}}(t_{21})\int\frac{d\omega}{2\pi}\Gamma_{\xi_{4}\xi_{3}\xi_{2}\xi_{1}}(\omega)\mathcal{I}_{\xi_{2}\xi_{1}}(\omega)\frac{e^{-i\omega t_{32}}[I_{\xi_{4}}(t_{43})-e^{-i\omega t_{43}}I_{\xi_{3}}^{*}(t_{43})]}{\omega-\mathcs{e}_{\xi_{4}}-\mathcs{e}_{\xi_{3}}+i(\gamma_{\xi_{4}}-\gamma_{\xi_{3}})}.\label{eqApp:TimeOrdRespFin30}
 \end{eqnarray}
Equations
(\ref{eqApp:TimeOrdRespFin10})-(\ref{eqApp:TimeOrdRespFin30})
constitute our most general expressions for the time domain nonlocal third order
responses in momentum space. The wavevectors and times in these
expressions correspond to the interaction events with the fields.
All three response functions depend only on time delays betwee different interactions.

We next turn to on the three signals defined by eqs. (\ref{eq:ThirdResponseRWA1})-(\ref{eq:ThirdResponseRWA3}).
The polarization evolution is
commonly transformed to the frequency-domain where the signal spectra are
displayed. The technique $\mathbf{k}_{I}$ is known as
photon echo. According to the Feynman diagrams (figure
\ref{enu:Comparison}) the system is transferred to a coherence after
the first interaction. The second interaction leaves the system
either in a population or in a coherence between two excitonic
states. We hold the second delay time, $T_2$ (often
referred to as population or waiting time) fixed. Then the
population and coherence evolution can be probed. The third
interaction creates coherences either between the ground and
one-exciton states or between one- and two-exciton states. To
display the signal we perform a double Fourier transform with
respect to the first and third time delays:
$T_{1}\rightarrow\Omega_{1}$ and $T_{3}\rightarrow\Omega_{3}$ in eqs. (\ref{eq:ThirdResponseRWA1})-(\ref{eq:ThirdResponseRWA3}). The
signal also depends on the directions of optical wavevectors. Thus,
the Fourier transform of eq. (\ref{eq:ThirdResponseRWA1}) can be
performed analytically since the Green's functions given by eq.
(\ref{eq:GreensEigen}) are simple exponential functions. We finally
obtain
\begin{equation}
\mathbs{P}_{\nu_{4}}^{\mathbf{k}_{I}}(\Omega_{3},T_{2},\Omega_{1})=\frac{1}{2^{3}}\exp(+i\phi_{3}+i\phi_{2}-i\phi_{1})\sum_{\nu_{3}\nu_{2}\nu_{1}} \mathbb{S}_{\nu_{4}\nu_{3}\nu_{2}\nu_{1}}^{\mathbf{k}_{I}}(\Omega_{3},T_2,\Omega_{1})E_{\nu_{3}}^{(3)}E_{\nu_{2}}^{(2)}E_{\nu_{1}}^{(1)}.\label{eq:P1Signal}
\end{equation}
In terms of the scattering matrix and the one-exciton
Green's functions, eq. (\ref{eqApp:TimeOrdRespFin10}) gives:
\begin{eqnarray}
 &  & \mathbb{S}_{\nu_{4}\nu_{3}\nu_{2}\nu_{1}}^{\mathbf{k}_{I}}(\Omega_{3},T_2,\Omega_{1})=
 2i\sum_{\xi_{4}...\xi_{1}}\langle\mathbs{d}_{\xi_{4}}^{\nu_{4}}(\mathbf{k}_{1}-\mathbf{k}_{2}-\mathbf{k}_{3})\mathbs{d}_{\xi_{3}}^{\nu_{3}\ast}(-\mathbf{k}_{3})\mathbs{d}_{\xi_{2}}^{\nu_{2}\ast}(-\mathbf{k}_{2})\mathbs{d}_{\xi_{1}}^{\nu_{1}}(-\mathbf{k}_{1})\rangle\nonumber \\
 &  & \quad\quad\times I_{\xi_{1}}^{*}(T_2)I_{\xi_{2}}(T_2)I_{\xi_{1}}^{\ast}(-\Omega_{1})I_{\xi_{4}}(\Omega_{3})
 \int_{-\infty}^{\infty}\frac{d\omega}{2\pi}\Gamma_{\xi_{4}\xi_{1}\xi_{3}\xi_{2}}(\omega)\mathcal{I}_{\xi_{3}\xi_{2}}(\omega)I_{\xi_{1}}^{\ast}(\omega-\Omega_{3}).\label{eq:SignalK1}
\end{eqnarray}
The frequency integration can be calculated analytically as follows. We note that
$\Gamma(\omega)\mathcal{I}(\omega)
    \sim
	    i(\omega - 2\bar{\mathcs{e}}+2i\bar{\gamma})^{-1}$
is a two-exciton Green's function with poles in negative immaginary half-plane (bars indicate averages), while $I_{\xi_{1}}^{\ast}(\omega-\Omega_{3})
=
-i(\omega-\Omega_{3} - \mathcs{e}_{\xi_{1}}-i\gamma_{\xi_{1}})^{-1}$
has one pole in positive immaginary half-plane:
$\omega_{p}=\Omega_{3} + \mathcs{e}_{\xi_{1}}+i\gamma_{\xi_{1}}$. In this case we can use the Cauchy integral formula by adding integration contour at the positive imaginary half-plane. Then
\begin{eqnarray}
 &  & \mathbb{S}_{\nu_{4}\nu_{3}\nu_{2}\nu_{1}}^{\mathbf{k}_{I}}(\Omega_{3},T_{2},\Omega_{1})=2i\sum_{\xi_{4}...\xi_{1}}\langle\mathbs{d}_{\xi_{4}}^{\nu_{4}}(\mathbf{k}_{1}-\mathbf{k}_{2}-\mathbf{k}_{3})\mathbs{d}_{\xi_{3}}^{\nu_{3}\ast}(-\mathbf{k}_{3})\mathbs{d}_{\xi_{2}}^{\nu_{2}\ast}(-\mathbf{k}_{2})\mathbs{d}_{\xi_{1}}^{\nu_{1}}(-\mathbf{k}_{1})\rangle\nonumber \\
 &  & \quad\quad\times I_{\xi_{1}}^{*}(T_2)I_{\xi_{2}}(T_2)I_{\xi_{1}}^{\ast}(-\Omega_{1})I_{\xi_{4}}(\Omega_{3})
\Gamma_{\xi_{4}\xi_{1}\xi_{3}\xi_{2}}(\Omega_{3} + \mathcs{e}_{\xi_{1}}+i\gamma_{\xi_{1}})\mathcal{I}_{\xi_{3}\xi_{2}}(\Omega_{3} + \mathcs{e}_{\xi_{1}}+i\gamma_{\xi_{1}}).\label{eq:SignalIntK1}
\end{eqnarray}

Similar to $\mathbf{k}_{I}$, $\mathbf{k}_{II}$ (which does not show an echo) can be defined with
the same Fourier transform and the same time delays:\begin{equation}
P_{\nu_{4}}^{\mathbf{k}_{II}}(\Omega_{3},T_2,\Omega_{1})=\frac{1}{2^{3}}\exp(+i\phi_{3}-i\phi_{2}+i\phi_{1})\sum_{\nu_{3}\nu_{2}\nu_{1}}
\mathbb{S}_{\nu_{4}\nu_{3}\nu_{2}\nu_{1}}^{\mathbf{k}_{II}}(\Omega_{3},T_2,\Omega_{1})E_{\nu_{3}}^{(3)}E_{\nu_{2}}^{(2)}E_{\nu_{1}}^{(1)}.\label{eq:P2Signal}
\end{equation}
In terms of the scattering matrix and the Green's functions we get from eq. (\ref{eqApp:TimeOrdRespFin20}):
\begin{eqnarray}
 &  & \mathbb{S}_{\nu_{4}\nu_{3}\nu_{2}\nu_{1}}^{\mathbf{k}_{II}}(\Omega_{3},T_2,\Omega_{1})=2i\sum_{\xi_{4}...\xi_{1}}\langle\mathbs{d}_{\xi_{4}}^{\nu_{4}}(\mathbf{k}_{2}-\mathbf{k}_{3}-\mathbf{k}_{1})\mathbs{d}_{\xi_{3}}^{\nu_{3}\ast}(-\mathbf{k}_{3})\mathbs{d}_{\xi_{2}}^{\nu_{2}}(-\mathbf{k}_{2})\mathbs{d}_{\xi_{1}}^{\nu_{1}\ast}(-\mathbf{k}_{1})\rangle\nonumber \\
 &  & \quad\quad\times I_{\xi_{2}}^{*}(T_2)I_{\xi_{1}}(T_2)I_{\xi_{1}}(\Omega_{1})I_{\xi_{4}}(\Omega_{3})\int_{-\infty}^{\infty}\frac{d\omega}{2\pi}\Gamma_{\xi_{4}\xi_{2},\xi_{3}\xi_{1}}(\omega)\mathcal{I}_{\xi_{3}\xi_{1}}(\omega)I_{\xi_{2}}^{\ast}(\omega-\Omega_{3})\label{eq:SignalK2}
\end{eqnarray}
and after integration over frequency
\begin{eqnarray}
 &  & \mathbb{S}_{\nu_{4}\nu_{3}\nu_{2}\nu_{1}}^{\mathbf{k}_{II}}(\Omega_{3},T_2,\Omega_{1})=
 2i\sum_{\xi_{4}...\xi_{1}}\langle\mathbs{d}_{\xi_{4}}^{\nu_{4}}(\mathbf{k}_{2}-\mathbf{k}_{3}-\mathbf{k}_{1})\mathbs{d}_{\xi_{3}}^{\nu_{3}\ast}(-\mathbf{k}_{3})\mathbs{d}_{\xi_{2}}^{\nu_{2}}(-\mathbf{k}_{2})\mathbs{d}_{\xi_{1}}^{\nu_{1}\ast}(-\mathbf{k}_{1})\rangle\nonumber \\
 &  & \quad\quad\times 
 I_{\xi_{2}}^{*}(T_2)
 I_{\xi_{1}}(T_2)
 I_{\xi_{1}}(\Omega_{1})
 I_{\xi_{4}}(\Omega_{3})
 \Gamma_{\xi_{4}\xi_{2},\xi_{3}\xi_{1}}(\Omega_3+\mathcs{e}_{\xi_{2}}+i\gamma_{\xi_{2}})
 \mathcal{I}_{\xi_{3}\xi_{1}}(\Omega_3+\mathcs{e}_{\xi_{2}}+i\gamma_{\xi_{2}}).\label{eq:SignalIntK2}
\end{eqnarray}
This expression is very similar to $\mathbf{k}_{I}$.

For $\mathbf{k}_{III}$, two interactions with the delay
$T_1$ create two-exciton coherence with
$\mathbf{k}_{1}+\mathbf{k}_{2}$. By performing the Fourier
transforms with respect to the second and third time delays:
$T_2\rightarrow\Omega_{2}$ and $T_3\rightarrow\Omega_{3}$ we
can observe the two-exciton coherence along $\Omega_{2}$ and mixed,
$0$-to-one and one-to-two, coherences along $\Omega_{3}$:
\begin{equation}
P_{\nu_{4}}^{\mathbf{k}_{III}}(\Omega_{3},\Omega_{2},T_1)=\frac{1}{2^{3}}\exp(-i\phi_{3}+i\phi_{2}+i\phi_{1})\sum_{\nu_{3}\nu_{2}\nu_{1}}\mathbb{S}_{\nu_{4}\nu_{3}\nu_{2}\nu_{1}}^{\mathbf{k}_{III}}(\Omega_{3},\Omega_{2},T_1)E_{\nu_{3}}^{(3)}E_{\nu_{2}}^{(2)}E_{\nu_{1}}^{(1)},\label{eq:P3Signal}
\end{equation}
which is obtained from eq. (\ref{eqApp:TimeOrdRespFin30}) as
\begin{eqnarray}
 &  & \mathbb{S}_{\nu_{4}\nu_{3}\nu_{2}\nu_{1}}^{\mathbf{k}_{III}}(\Omega_{3},\Omega_{2},T_1)=2i\sum_{\xi_{4}...\xi_{1}}\langle\mathbs{d}_{\xi_{4}}^{\nu_{4}}(\mathbf{k}_{3}-\mathbf{k}_{2}-\mathbf{k}_{1})\mathbs{d}_{\xi_{3}}^{\nu_{3}}(-\mathbf{k}_{3})\mathbs{d}_{\xi_{2}}^{\nu_{2}\ast}(-\mathbf{k}_{2})\mathbs{d}_{\xi_{1}}^{\nu_{1}\ast}(-\mathbf{k}_{1})\rangle\nonumber \\
 &  & \quad\quad\times 
 I_{\xi_{1}}(T_1)
 I_{\xi_{4}}(\Omega_{3})
 \int_{-\infty}^{\infty}\frac{d\omega}{2\pi}
 \Gamma_{\xi_{4}\xi_{3},\xi_{2}\xi_{1}}(\omega)\mathcal{I}_{\xi_{2}\xi_{1}}(\omega)
 I_{\xi_{3}}^{\ast}(\omega-\Omega_{3})\theta(\Omega_{2}-\omega).\label{eq:SignalK3}
\end{eqnarray}
 The function $\theta(\omega)=i(\omega+i\gamma')^{-1}$ is taken in
the limit $\gamma_{\xi}\gg\gamma'>0$.
Analytic integration over frequency  involves more terms. Again,
$\Gamma(\omega)\mathcal{I}(\omega)
    \sim
	    i(\omega - 2\bar{\mathcs{e}}+2i\bar{\gamma})^{-1}$
is a two-exciton Green's function with poles in the negative imaginary half-plane, while $I_{\xi_{1}}^{\ast}(\omega-\Omega_{3})
=
-i(\omega-\Omega_{3} - \mathcs{e}_{\xi_{1}}-i\gamma_{\xi_{1}})^{-1}$
has one pole in positive immaginary half-plane:
$\omega_{p}=\Omega_{3} + \mathcs{e}_{\xi_{1}}+i\gamma_{\xi_{1}}$.
$\theta(\Omega_2 - \omega)=-i(\omega-\Omega_2 - i\gamma')^{-1}$ has another pole with positive immaginary part: $\omega_{p_2}=\Omega_2 + i\gamma'$.
In this case Cauchy integration results into two terms and
\begin{eqnarray}
 &  & \mathbb{S}_{\nu_{4}\nu_{3}\nu_{2}\nu_{1}}^{\mathbf{k}_{III}}(\Omega_{3},\Omega_{2},T_1)=2i\sum_{\xi_{4}...\xi_{1}}\langle\mathbs{d}_{\xi_{4}}^{\nu_{4}}(\mathbf{k}_{3}-\mathbf{k}_{2}-\mathbf{k}_{1})\mathbs{d}_{\xi_{3}}^{\nu_{3}}(-\mathbf{k}_{3})\mathbs{d}_{\xi_{2}}^{\nu_{2}\ast}(-\mathbf{k}_{2})\mathbs{d}_{\xi_{1}}^{\nu_{1}\ast}(-\mathbf{k}_{1})\rangle\nonumber \\
 &  & \quad\quad\times 
 I_{\xi_{1}}(T_1)
 I_{\xi_{4}}(\Omega_{3})
 I_{\xi_{3}}^{\ast}(\Omega_2-\Omega_{3})
  [
 \Gamma_{\xi_{4}\xi_{3},\xi_{2}\xi_{1}}(\Omega_2)\mathcal{I}_{\xi_{2}\xi_{1}}(\Omega_2) \nonumber \\
 &  & \quad\quad \quad\quad - \Gamma_{\xi_{4}\xi_{3},\xi_{2}\xi_{1}}(\Omega_3+\mathcs{e}_{\xi_3}+i\gamma_{\xi_3})\mathcal{I}_{\xi_{2}\xi_{1}}(\Omega_3+\mathcs{e}_{\xi_3}+i\gamma_{\xi_3})],\label{eq:SignalIntK3}
\end{eqnarray}
where we have used $\gamma_{\xi} \gg \gamma'$.

\section{The sequential pump-probe signal}

\label{sec:PPSignal}

Sequential pump-probe is an incoherent two-pulse FWM technique
commonly used for monitoring excited state dynamics. The pump and the probe
pulses are assumed well-separated and characterized by their carrier
frequencies $\omega_{pu}$ for pump and $\omega_{pr}$ for probe, and
their delay $\tau_{pp}$. We ignore population transport and assume
that the delay between pump and probe shorter than the population
evolution time. In addition, we assume that the pulse bandwidths are
much narrower than the exciton bandwidth (which always holds for
electronic spectroscopy).

There are two interactions with the pump pulse
($\mathbf{k}_{1}=\pm\mathbf{k}_{pu}$,
$\mathbf{k}_{2}=\mp\mathbf{k}_{pu}$, where $\mathbf{k}_{pu}$ is the
pump pulse wavevector) and one with the probe
($\mathbf{k}_{3}=\pm\mathbf{k}_{pr}$, where $\mathbf{k}_{pr}$ is the
probe wavevector). We assume that the pump has various polarization
components with phases $\phi_{pu}^{\nu}$ in eq.
(\ref{eq:ActingOptField}), which can describe e.g. circular
polarization of pump. The probe can also have all polarization
components with phases $\phi_{pr}^{\nu}$. The pump-probe
experiment measures change of absorption of probe with respect to
its linear absorption. The absorption of the optical field is given by
\cite{mukbook,DomckeStock97Rev}:
\begin{eqnarray}
\sigma_{A} & = & \int d\mathbf{x}[\frac{\partial\mathbs{P}_{\nu}(\mathbf{x})}{\partial t}\mathbs{E}_{\nu}(\mathbf{x})],
\end{eqnarray}
 where $\mathbs{P}_{\nu}(\mathbf{x})$ is a third order induced polarization and $\mathbs{E}_{\nu}(\mathbf{x})$ is a probe field
when considering pump-probe experiment. Using the response function
(eq. (\ref{eq:ThirdResponse})) we obtain for the pump-probe
signal:
\begin{eqnarray}
 &  & \sigma_{PP}(\omega_{pr},\tau_{pp},\omega_{pu})=\frac{1}{16}\sum_{\nu_{4}\nu_{3}\nu_{2}\nu_{1}}\int dt_{4}\int dt_{3}\int dt_{2}\int dt_{1}\nonumber \\
 &  & \qquad\tilde{E}_{\nu_{4}}^{(pr)}(t_{4}-t_{pr})\tilde{E}_{\nu_{3}}^{(pr)}(t_{3}-t_{pr})\tilde{E}_{\nu_{2}}^{(pu)}(t_{2}-t_{pu})\tilde{E}_{\nu_{1}}^{(pu)}(t_{1}-t_{pu})\nonumber \\
 &  & \qquad\times\Big\{\frac{\partial S_{\nu_{4}\nu_{3}\nu_{2}\nu_{1}}^{(3)}(-\mathbf{k}_{pr}t_{4},+\mathbf{k}_{pr}t_{3},+\mathbf{k}_{pu}t_{2},-\mathbf{k}_{pu}t_{1})}{\partial t_{4}}\nonumber \\
 &  & \qquad\times\exp[i\omega_{pr}(t_{4}-t_{3})-i\omega_{pu}(t_{2}-t_{1})+i(-\phi_{pr}^{\nu_{4}}+\phi_{pr}^{\nu_{3}}+\phi_{pu}^{\nu_{2}}-\phi_{pu}^{\nu_{1}})]\nonumber \\
 &  & \qquad+\frac{\partial S_{\nu_{4}\nu_{3}\nu_{2}\nu_{1}}^{(3)}(-\mathbf{k}_{pr}t_{4},+\mathbf{k}_{pr}t_{3},-\mathbf{k}_{pu}t_{2},+\mathbf{k}_{pu}t_{1})}{\partial t_{4}}\nonumber \\
 &  & \qquad\times\exp[i\omega_{pr}(t_{4}-t_{3})+i\omega_{pu}(t_{2}-t_{1})+i(-\phi_{pr}^{\nu_{4}}+\phi_{pr}^{\nu_{3}}-\phi_{pu}^{\nu_{2}}+\phi_{pu}^{\nu_{1}})]\nonumber \\
 &  & \qquad+\frac{\partial S_{\nu_{4}\nu_{3}\nu_{2}\nu_{1}}^{(3)}(+\mathbf{k}_{pr}t_{4},-\mathbf{k}_{pr}t_{3},+\mathbf{k}_{pu}t_{2},-\mathbf{k}_{pu}t_{1})}{\partial t_{4}}\nonumber \\
 &  & \qquad\times\exp[-i\omega_{pr}(t_{4}-t_{3})-i\omega_{pu}(t_{2}-t_{1})+i(+\phi_{pr}^{\nu_{4}}-\phi_{pr}^{\nu_{3}}+\phi_{pu}^{\nu_{2}}-\phi_{pu}^{\nu_{1}})]\nonumber \\
 &  & \qquad+\frac{\partial S_{\nu_{4}\nu_{3}\nu_{2}\nu_{1}}^{(3)}(+\mathbf{k}_{pr}t_{4},-\mathbf{k}_{pr}t_{3},-\mathbf{k}_{pu}t_{2},+\mathbf{k}_{pu}t_{1})}{\partial t_{4}}\nonumber \\
 &  & \qquad\times\exp[-i\omega_{pr}(t_{4}-t_{3})+i\omega_{pu}(t_{2}-t_{1})+i(+\phi_{pr}^{\nu_{4}}-\phi_{pr}^{\nu_{3}}-\phi_{pu}^{\nu_{2}}+\phi_{pu}^{\nu_{1}})]\Big\}
\end{eqnarray}
Invoking the RWA and the other assumptions regarding the pulses made
in section \ref{sec:NonLinearAbsorption}, we find that this signal
is a combination of the $\mathbf{k}_{I}$ and $\mathbf{k}_{II}$
techniques:
\begin{eqnarray}
 &  & \sigma_{PP}(\omega_{pr},\tau_{pp},\omega_{pu})=\frac{1}{8}\omega_{pr}\sum_{\nu_{4}\nu_{3}\nu_{2}\nu_{1}}E_{\nu_{4}}^{(pr)}E_{\nu_{4}}^{(pr)}E_{\nu_{4}}^{(pu)}E_{\nu_{4}}^{(pu)}\nonumber \\
 &  & \qquad\times Im \Big\{ \mathbb{S}_{\nu_{4}\nu_{3}\nu_{2}\nu_{1}}^{\mathbf{k}_{I}}(\omega_{pr},\tau_{pp},-\omega_{pu})e^{i(-\phi_{pr}^{\nu_{4}}+\phi_{pr}^{\nu_{3}}+\phi_{pu}^{\nu_{2}}-\phi_{pu}^{\nu_{1}})}\nonumber \\
 &  & \qquad+\mathbb{S}_{\nu_{4}\nu_{3}\nu_{2}\nu_{1}}^{\mathbf{k}_{II}}(\omega_{pr},\tau_{pp},\omega_{pu})e^{i(-\phi_{pr}^{\nu_{4}}+\phi_{pr}^{\nu_{3}}-\phi_{pu}^{\nu_{2}}+\phi_{pu}^{\nu_{1}})}\Big\}\label{eq:PP-final}
\end{eqnarray}
where
$\mathbf{k}_{I}=\mathbf{k}_{pr}+\mathbf{k}_{pu}-\mathbf{k}_{pu}$ and
$\mathbf{k}_{II}=\mathbf{k}_{pr}-\mathbf{k}_{pu}+\mathbf{k}_{pu}$ .
It is possible to measure chiral components of this signal using
left and right handed circular polarizations so that non chiral
components cancel. When the pump bandwidth is larger than the
exciton bandwidth eq. (\ref{eq:PP-final}) needs to be integrated
over $\omega_{pu}$.

\section{Non-time-ordered response function and frequency-domain FWM signals}

\label{sec:susceptibilities}

Using eq. (\ref{eq:P3Complete}) we can define a non time ordered response function by:
\begin{eqnarray}
 &  & S_{\nu_{4}...\nu_{1}}(\mathbf{x}_{4},...,\mathbf{x}_{1})=
\frac{i}{3}\sum_{perm_3}\sum_{n_{4}...n_{1}}\langle\mathbf{M}_{n_{4}n_{3}n_{2}n_{1}}^{\nu_{4}\nu_{3}\nu_{2}\nu_{1}}(\mathbf{r}_{4},\mathbf{r}_{3},\mathbf{r}_{2},\mathbf{r}_{1})\rangle\nonumber \\
 &  & \quad\quad\times\int_{-\infty}^{\infty}d\tau''\int_{-\infty}^{\infty}d\tau'\sum_{n_{4}'n_{3}'n_{2}'n_{1}'}\Gamma_{n_{4}'n_{3}'n_{2}'n_{1}'}(\tau''-\tau').\nonumber \\
 &  & \quad\quad\times G_{n_{4}n_{4}'}(t_4-\tau'')G_{n_{3}'n_{3}}^{\dagger}(\tau''-t_3)G_{n_{2}'n_{2}}(\tau'-t_2)G_{n_{1}'n_{1}}(\tau'-t_1)+c.c.,\label{eq:TUOrd3}
 \end{eqnarray}
where $\sum_{perm_3}$ denotes three terms in the following permutation:
$(\nu_3\mathbf{r}_3t_3,\nu_2\mathbf{r}_2t_2,\nu_1\mathbf{r}_1t_1)$,
$(\nu_2\mathbf{r}_2t_2,\nu_3\mathbf{r}_3t_3,\nu_1\mathbf{r}_1t_1)$
and
$(\nu_1\mathbf{r}_1t_1,\nu_2\mathbf{r}_2t_2,\nu_3\mathbf{r}_3t_3)$.
This form can be also expressed in the eigenstate basis using frequency domain scattering matrix.

Eq. (\ref{eq:TUOrd3}) can be substituted into eq. (\ref{eq:ThirdResponse}) to give a non time ordered representation of the response. This form is very convenient for frequency domain four-wave-mixing processes \cite{mukbook}: \begin{eqnarray}
 &  & \mathbs{P}_{\nu_{4}}^{(3)}(\mathbf{k}_{4},\omega_{4})=\frac{1}{(2\pi)^{12}}\sum_{\nu_{3}\nu_{2}\nu_{1}}\int d\mathbf{k}_{3}\int d\omega_{3}\int d\mathbf{k}_{2}\int d\omega_{2}\int d\mathbf{k}_{1}\int d\omega_{1}\times\nonumber \\
 &  & \qquad\qquad\chi_{\nu_{4},\nu_{3}\nu_{2}\nu_{1}}^{(3)}(-\mathbf{k}_{4}-\omega_{4};\mathbf{k}_{3}\omega_{3},\mathbf{k}_{2}\omega_{2},\mathbf{k}_{1}\omega_{1})\mathbs{E}_{\nu_{3}}(\mathbf{k}_{3},\omega_{3})\mathbs{E}_{\nu_{2}}(\mathbf{k}_{2},\omega_{2})\mathbs{E}_{\nu_{1}}(\mathbf{k}_{1},\omega_{1})\label{eq:3PolarizationW},\end{eqnarray}
where the susceptibility is obtained from eq. (\ref{eq:TUOrd3}) \cite{Abramavicius2005jcp-chiral}:
\begin{eqnarray}
 &  & \chi_{\nu_{4},\nu_{3}\nu_{2}\nu_{1}}^{(3)}(-\mathbf{k}_{4},-\omega_{4};\mathbf{k}_{3},\omega_{3},\mathbf{k}_{2},\omega_{2},\mathbf{k}_{1},\omega_{1})=2\pi i\delta(\omega_{4}-\omega_{3}-\omega_{2}-\omega_{1})\nonumber \\
 &  & \quad\quad
\times \frac{1}{3}
 \sum_{perm_3}\sum_{\xi_{4}\xi_{3}\xi_{2}\xi_{1}}\langle\mathbs{d}_{\xi_{4}}^{\nu_{4}}(\mathbf{k}_{4})\mathbs{d}_{\xi_{3}}^{\nu_{3}}(-\mathbf{k}_{3})\mathbs{d}_{\xi_{2}}^{\nu_{2}\ast}(\mathbf{k}_{2})\mathbs{d}_{\xi_{1}}^{\nu_{1}\ast}(\mathbf{k}_{1})\rangle\nonumber \\
 &  & \quad\quad \times
 \Gamma_{\xi_{4}\xi_{3},\xi_{2}\xi_{1}}(\omega_{2}+\omega_{1})I_{\xi_{4}}(\omega_{4})I_{\xi_{3}}^{\ast}(-\omega_{3})I_{\xi_{2}}(\omega_{2})I_{\xi_{1}}(\omega_{1})+c.c.'.\label{eq:ResponseWExcitonAv}
 \end{eqnarray}
where $\sum_{perm_3}$ denotes a sum over the three permutations: $(\nu_3\mathbf{k}_3\omega_3,\nu_2\mathbf{k}_2\omega_2,\nu_1\mathbf{k}_1\omega_1)$,
$(\nu_2\mathbf{k}_2\omega_2,\nu_3\mathbf{k}_3\omega_3,\nu_1\mathbf{k}_1\omega_1)$
and
$(\nu_1\mathbf{k}_1\omega_1,\nu_2\mathbf{k}_2\omega_2,\nu_3\mathbf{k}_3\omega_3)$ (the expression is already symmetric to the permutation of $\nu_1\mathbf{k}_1\omega_1$ and $\nu_2\mathbf{k}_2\omega_2$).

We consider CW laser fields characterized by their amplitude $E_{\nu}^{(s)}$, wavevector $\mathbf{k}_s$ and optical frequency $\omega_s$:
\begin{equation}
\mathbs{E}_{\nu}(\mathbf{k},\omega) = \frac{(2\pi)^4}{2}
E_{\nu}^{(s)}e^{i\phi_s^{\nu}}\delta(\mathbf{k}+\mathbf{k}_s)\delta(\omega-\omega_s)+ c.c.'
\end{equation}
We consider the signal in the direction
 $-\mathbf{k}_1-\mathbf{k}_2+\mathbf{k}_3$:
\begin{eqnarray}
&  & \mathbs{P}_{\nu_{4}}^{(3)}(-\mathbf{k}_1-\mathbf{k}_2+\mathbf{k}_3,\omega_{1}+\omega_2-\omega_3)
=\frac{1}{2^{3}}\sum_{\nu_{3}\nu_{2}\nu_{1}}e^{-i\phi_3^{\nu_3}+i\phi_2^{\nu_2}+i\phi_1^{\nu_1}}
 E_{\nu_{3}}^{(3)}
 E_{\nu_{2}}^{(2)}
 E_{\nu_{1}}^{(1)}
  \nonumber \\
&  & \qquad\qquad
\times
\chi_{\nu_{4},\nu_{3}\nu_{2}\nu_{1}}^{(3)}(\mathbf{k}_1+\mathbf{k}_2-\mathbf{k}_3,-\omega_{1}-\omega_2+\omega_3;\mathbf{k}_{3},-\omega_{3},-\mathbf{k}_{2}\omega_{2},-\mathbf{k}_{1}\omega_{1})
 \label{eq:3PolarizationW11},\end{eqnarray}
In the RWA this gives
\begin{eqnarray}
 &  &
\chi_{\nu_{4},\nu_{3}\nu_{2}\nu_{1}}^{(3)}(\mathbf{k}_{1}+\mathbf{k}_{2}-\mathbf{k}_{3},-\omega_{1}-\omega_{2}+\omega_{3};\mathbf{k}_{3},-\omega_{3},-\mathbf{k}_{2},\omega_{2},-\mathbf{k}_{1},\omega_{1})=2\pi i\nonumber \\
 &  & \quad\quad
\times \frac{1}{3}
\sum_{\xi_{4}\xi_{3}\xi_{2}\xi_{1}}\langle\mathbs{d}_{\xi_{4}}^{\nu_{4}}(-\mathbf{k}_{1}-\mathbf{k}_{2}+\mathbf{k}_{3})\mathbs{d}_{\xi_{3}}^{\nu_{3}}(-\mathbf{k}_{3})\mathbs{d}_{\xi_{2}}^{\nu_{2}\ast}(-\mathbf{k}_{2})\mathbs{d}_{\xi_{1}}^{\nu_{1}\ast}(-\mathbf{k}_{1})\rangle\nonumber \\
 &  & \quad\quad \times
\Gamma_{\xi_{4}\xi_{3},\xi_{2}\xi_{1}}(\omega_{2}+\omega_{1})I_{\xi_{4}}(\omega_{1}+\omega_{2}-\omega_{3})I_{\xi_{3}}^{\ast}(\omega_{3})I_{\xi_{2}}(\omega_{2})I_{\xi_{1}}(\omega_{1})+c.c.'.\label{eq:ResponseWExcitonAv_}
 \end{eqnarray}
This is a three dimensional signal of optical frequencies. Two dimensional sections at $\omega_3 = \omega_1$ of that signal can be used \cite{Abramavicius2005jcp-chiral}.

\section{The exciton scattering matrix for periodic structures}

\label{sec:The-exciton-infinite}

In this appendix we calculate the time domain scattering matrix (eq.
(\ref{eq:FunctionJ})) for periodic systems, which can be easily
extended to non periodic finite systems. Using eq.
(\ref{eq:ScattMatrixW}) we calculate the scattering matrix in the
frequency domain, which is directly used in appendix
\ref{sec:TDSignals}.  The time domain scattering matrix is then
given by the Fourier transform:
\begin{eqnarray}
 && \Gamma_{\xi_{4}\xi_{3}\xi_{2}\xi_{1}}(t) = \int d\omega \exp(-i\omega t) \Gamma_{\xi_{4}\xi_{3}\xi_{2}\xi_{1}}(\omega) \label{eq:ScatRelationTW}
\end{eqnarray}

An infinite system is constructed by replicating a unit cell where
all vibrational or electronic modes are fixed at particular sites
inside each cell in some crystal lattice. We will assume a cubic
lattice of dimensionality $\mathcal{D}$ with a finite number
$N^{\mathcal{D}}$ cells. To guarantee translational invariance we
use cyclic boundary conditions. Each mode is represented by a pair
of indices, $\mathbf{R}m$, where $\mathbf{R}$ is a position vector
of the cell and $m$ is the index of the site within the unit cell;
the position of the $m$-th site inside the molecule is given by the
vector $\mathbf{R}+\mathbs{\rho}_{m}$ where $\mathbf{R}$ is the
origin of the unit cell and $\mathbs{\rho}_{m}$ is the displacement
from that origin. We denote the number of sites in the cell by
$\mathcal{M}$ and the lattice constant $a$. Since the system is
translationally invariant, the intermode coupling
$J_{\mathbf{R}m,\mathbf{R}'n}=J_{m,n}(\mathbf{R}'-\mathbf{R})$ now
depends on the distance between cells $\mathbf{R}'-\mathbf{R}$ and
on the site indices of each cell, $m$ and $n$. When
$\mathbf{R}'=\mathbf{R}$, $J_{m,n}(0)$ describes the coupling of
modes inside the cell. $J_{m,n}(\mathbf{R}'-\mathbf{R})$ with
$\mathbf{R}'\neq\mathbf{R}$ defines inter-cell couplings. Similar to
this coupling the intermode anharmonicity
$\Delta_{\mathbf{R}m,\mathbf{R}'n}=\Delta_{m,n}(\mathbf{R}'-\mathbf{R})$
now depends on the distance between cells and on the site indices of
each cell. We note that $\Delta_{m,m}(\mathbf{R}'-\mathbf{R})$ with
$\mathbf{R}'\neq\mathbf{R}$ defines anharmonicity of the combination
band, where two excitations are localized on different sites.

The one-exciton eigenstates $\xi$ of a periodic system are
characterized by two quantum numbers: the Davydov band index
$\lambda$ is related to different sites in the unit cell, and the
exciton momentum (Bloch wavevector) $\mathbf{q}$.  For molecular
systems much smaller than the optical wavelength only zero momentum,
 $\mathbf{q}=0$, exciton states contribute to nonlinear optical
response. The infinite size only enters into the scattering matrix,
where excitons with different momenta can be involved in the exciton
scattering process.

The equations for the optical response (\ref{eq:SignalK1}) --
(\ref{eq:SignalK3}) can be equivalently used for periodic structures
with cyclic boundary conditions provided the eigenstates $\xi$ are
replaced with the periodic system eigenstates $\lambda$ at momentum
$\mathbf{q}=0$. The scattering matrix is then considerably
simplified:
\begin{eqnarray}
 &  & \Gamma_{\lambda_{4}\lambda_{3}\lambda_{2}\lambda_{1}}(\omega)={\sum_{m_{4}...m_{1}}}'\bar{\psi}_{\lambda_{4}m_{4}}\bar{\psi}_{\lambda_{3}m_{3}}\big[\sum_{-l_{c}<\mathbf{r}'\mathbf{r}''<l_{c}}\bar{\Gamma}_{\mathbf{r}''m_{4}m_{3};\mathbf{r}'m_{2}m_{1}}(\omega)\big]\bar{\psi}_{\lambda_{2}m_{2}}\bar{\psi}_{\lambda_{1}m_{1}},\label{eq:ScatMatrixFinal}
\end{eqnarray}
where $\bar{\Gamma}_{\mathbf{r}_{1},m,n;\mathbf{r}_{2},m',n'}(\omega)$
is the mixed space scattering matrix (taken at zero momentum).
This scattering matrix
is given in momentum space with respect to the translational motion of
excitons, while it explicitly depends on real space coordinates $\mathbf{r}_1$ and $\mathbf{r}_2$,
which are the distances between pairs of cells  -- these are not translationally invariant.
This scattering matrix is given by \cite{Abramavicius2005jcp-chiral}:
\begin{eqnarray}
 &  & \bar{\Gamma}_{\mathbf{r}_{1},m,n;\mathbf{r}_{2},m',n'}(\omega)=-i\Delta_{mn}(\mathbf{r}_{1})(\bar{D}(\omega))_{\mathbf{r}_{1},m,n;\mathbf{r}_{2},m',n'}^{-1}\label{eq:AppRedScatMatrixFinal}
 \end{eqnarray}
 and the matrix
\begin{eqnarray}
 &  & \bar{D}_{\mathbf{r}_{1},m,n;\mathbf{r}_{2},m',n'}(\omega)\nonumber \\
 &  & \quad\quad=\delta_{\mathbf{r}_{1},\mathbf{r}_{2}}\delta_{m,m'}\delta_{n,n'}+i\bar{\mathcal{G}}_{\mathbf{r}_{1},m,n;\mathbf{r}_{2},m',n'}(\omega)\Delta_{m'n'}(\mathbf{r}_{2}),\label{eqApp:BarDAt0}
\end{eqnarray}
 where the mixed space two-exciton Green's function
\begin{equation}
\bar{\mathcal{G}}_{\mathbf{r}_{1},m,n;\mathbf{r}_{2},m',n'}(\omega)=\frac{1}{\mathcal{V}}\sum_{\mathbf{q}}e^{-i\mathbf{q}(\mathbf{r}_{2}-\mathbf{r}_{1})}g_{m,n;m',n'}(\mathbf{q},-\mathbf{q},\omega),\label{eqApp:Greens2At0}
\end{equation}
 involves the sum over two-exciton Green's function of one unit cell
with different momenta:
\begin{equation}
g_{m,n;m',n'}(\mathbf{q},-\mathbf{q},\omega)={\sum_{\lambda\lambda'}}'\bar{\psi}_{\lambda m}(\mathbf{q})\bar{\psi}_{\lambda'n}(-\mathbf{q})\mathcal{I}_{\lambda\lambda'}(\mathbf{q},-\mathbf{q},\omega)\bar{\psi}_{\lambda m'}^{*}(\mathbf{q})\bar{\psi}_{\lambda'n'}^{*}(-\mathbf{q}).\label{eqApp:Greens2OfCellRed}
\end{equation}
We have adopted the following notation. $\bar{\psi}_{\lambda m}$ is
a zero momentum wavefunction of exciton band $\lambda$ obtained from
${\sum_{m'}}'J_{m,m'}(\mathbf{q}=0)\bar{\psi}_{\lambda
m'}=\mathcs{e}_{\lambda}\bar{\psi}_{\lambda m}$, where
$J_{m,m'}(\mathbf{q}=0)=\sum_{\mathbf{r}}J_{m,m'}(\mathbf{r})$ and
$\mathcs{e}_{\lambda}$ is the eigenenergy of this state. The prime in
the sums denotes the summation either over sites within one cell or
over different Davydov bands at zero momentum, while the sum over
$\mathbf{r}$ is a sum over cells including $\mathbf{r}=0$ within the
scattering length $l_{c}$ defined by
$\Delta_{mn}(\mathbf{r}>l_{c})=0$. All site indices $m$ run within
one cell, while $\sum_{\mathbf{q}}$ is the sum over momenta. For
infinite systems this sum becomes an integral over all exciton
bands.

The scattering matrix calculation (eq. (\ref{eq:FunctionJ})) is the
same for the finite non-periodic system except that $\mathbf{r}$ and
$\mathbf{q}$ are set to 0 in eqs.
(\ref{eq:ScatMatrixFinal})-(\ref{eqApp:Greens2OfCellRed}), site
indices $m$ and $n$ then correspond to different modes in the entire
system and the exciton band indices $\lambda$ are changed into
exciton states $\xi$.


\newpage

\input{tables.tex}

\newpage



\newpage

\renewcommand{\labelenumi}{Figure \arabic{enumi}.}

Figure Captions

\begin{enumerate}
\item \label{enu:Scheme}(top)
Time domain third order experiments:
three short laser pulses with wavevectors $\mathbf{k}_{1}$,
$\mathbf{k}_{2}$ and $\mathbf{k}_{3}$ generate a nonlinear
polarization with wavevector $\mathbf{k}_{S}$. (bottom left)
Structures of the peptide backbone in $\alpha$ helix and
antiparallel $\beta$ sheet structures: green - C atoms, blue - N
atoms, red - O atoms. C=O responsible for amide I mode (1600-1700
cm\sups{-1}) are emphasized. (bottom right) Energy level scheme of
excitonic system with one ground state, a manifold of one-exciton
states ($e$) and a manifold with two-exciton states ($f$).
\item \label{enu:Notations}
a) Time variables used in calculating the excitonic response. Red
peaks indicate laser pulses, blue area corresponds to the
exciton-exciton scattering process. $t_{1}$, $t_{2}$ and $t_{3}$ are
the first, second and third interactions with laser pulses. $t_{4}$
is the signal generation time. $t_{i+1,i} \equiv t_{i+1}-t_{i}$ are
the delay times (always positive) between two interactions.
$\tau_{s}''$ and $\tau_{s}'$ are the delay times of the exciton
scattering. b) Transformation from the non ordered time variables
$\tau_{1}$, $\tau_{2}$ and $\tau_{3}$ to ordered times $t_{1}$,
$t_{2}$ and $t_{3}$ in eqs.
(\ref{eq:P3Complete}-\ref{eq:TimeOrderedResponse}) which defines
the three different scattering pathways. c) Three scattering pathways:
$\mathbf{k}_{I}=-\mathbf{k}_{1}+\mathbf{k}_{2}+\mathbf{k}_{3}$ involves scattering
of the excitons created by $\mathbf{k}_{2}$ and $\mathbf{k}_{3}$,
$\mathbf{k}_{II}=\mathbf{k}_{1}-\mathbf{k}_{2}+\mathbf{k}_{3}$ involves scattering
of the excitons created by $\mathbf{k}_{1}$ and $\mathbf{k}_{3}$,
and $\mathbf{k}_{III}=\mathbf{k}_{1}+\mathbf{k}_{2}-\mathbf{k}_{3}$ involves
scattering of the excitons created by $\mathbf{k}_{1}$ and
$\mathbf{k}_{2}$.
\item \label{enu:SignalLA}
Linear absorption of the $\alpha$ helix (top)  and the antiparallel  $\beta$ sheet (bottom).
\item \label{enu:SignalK1}
Absolute value of $\mathbb{S}_{\nu_{4}\nu_{3}\nu_{2}\nu_{1}}^{\mathbf{k}_{I}}(\Omega_{3},T_2=0,\Omega_{1})$
signal of the $\alpha$ helix and antiparallel $\beta$ sheet (eq.
(\ref{eq:SignalK1})). Shown are one nonchiral, $xxxx$, and three
 chiral, $xxxy$, $xxyx$, $xyxx$, components as indicated. Blue crosses
mark the crosspeaks of $xxxx$.
\item \label{enu:SignalK2}
Absolute value of the $\mathbb{S}_{\nu_{4}\nu_{3}\nu_{2}\nu_{1}}^{\mathbf{k}_{III}}(\Omega_{3},\Omega_{2},T_1=0)$
signal of the $\alpha$ helix and antiparallel $\beta$ sheet (eq.
(\ref{eq:SignalK3})). Shown are one nonchiral, $xxxx$, and three
chiral, $xxxy$, $xxyx$, $xyxx$, components, as indicated.
\item \label{enu:Comparison}
Comparison of the exciton-scattering and
the transition-among-eigenstates pictures
of the  three signals $\mathbb{S}^{\mathbf{k}_{I}}(\Omega_{3},T_2,\Omega_{1})$,
$\mathbb{S}^{\mathbf{k}_{II}}(\Omega_{3},T_2,\Omega_{1})$ and
$\mathbb{S}^{\mathbf{k}_{III}}(\Omega_{3},\Omega_{2},T_1)$.
$\Omega_{1}$, $\Omega_{2}$ and $\Omega_{3}$ are the
Fourier transform variables conjugate to $T_1$, $T_2$ and $T_3$.
Red dots represent the interaction of the system with the field,
blue dots mark the exciton scattering space.
\end{enumerate}

\end{document}